\documentclass[twocolumn,pre,aps,showpacs,showkeys,amsmath]{revtex4}
\usepackage{graphicx}
\usepackage{adjustbox}
\usepackage{multirow}

\begin{document}

\title{Quantifying Harmony between Direct and Indirect Pathways in The Basal Ganglia; Healthy and Parkinsonian States}
\author{Sang-Yoon Kim}
\email{sykim@icn.re.kr}
\author{Woochang Lim}
\email{wclim@icn.re.kr}
\affiliation{Institute for Computational Neuroscience and Department of Science Education, Daegu National University of Education, Daegu 42411, Korea}

\begin{abstract}
The basal ganglia (BG) show a variety of functions for motor and cognition. There are two competitive pathways in the BG; direct pathway (DP) which facilitates movement and indirect pathway (IP) which suppresses movement. It is well known that diverse functions of the BG may be made through ``balance'' between DP and IP. But, to the best of our knowledge, so far no quantitative analysis for such balance was done. In this paper, as a first time, we introduce the competition degree
${\cal C}_d$ between DP and IP. Then, by employing ${\cal C}_d$, we quantify their competitive harmony (i.e., competition and cooperative interplay), which could lead to improving our understanding of the traditional ``balance'' so clearly and quantitatively. We first consider the case of normal dopamine (DA) level of $\phi^*=0.3$. In the case of phasic cortical input (10 Hz), a healthy state with ${\cal C}_d^* = 2.82$ (i.e., DP is 2.82 times stronger than IP) appears. In this case, normal movement occurs via harmony between DP and IP. Next, we consider the case of decreased DA level, $\phi = \phi^*(=0.3)~x_{DA}$ ($1 > x_{DA} \geq 0$). With decreasing $x_{DA}$ from 1, the competition degree ${\cal C}_d$ between DP and IP decreases monotonically from ${\cal C}_d^*$, which results in appearance of a pathological Parkinsonian state with reduced ${\cal C}_d$. In this Parkinsonian state, strength of IP is much increased than that in the case of normal healthy state, leading to disharmony between DP and IP. Due to such break-up of harmony between DP and IP, impaired movement occurs. Finally, we also study treatment of the pathological Parkinsonian state via recovery of harmony between DP and IP.
\end{abstract}

\pacs{87.19.lj, 87.19.lu, 87.19.rs}

\keywords{Quantitative analysis, Competition degree, Harmony, Direct pathway, Indirect pathway, Basal ganglia, Healthy state, Parkinsonian state}

\maketitle

\section{Introduction}
\label{sec:INT}
The basal ganglia (BG) in the brain are a group of subcortical deep-lying nuclei, take cortical inputs from most regions of cortex, and
make inhibitory output projection to the thalamus/brainstem \cite{Luo,Kandel,Squire,Bear}. Their main function is motor control (e.g., initiation and execution of movement) \cite{Luo,Kandel,Squire,Bear}. They also play an important role in cognitive processes (e.g., action selection) \cite{GPR1,GPR2,Hump1,Hump2,Hump3,Man}. Dysfunction in the BG is associated with a number of movement disorders, such as Parkinson's disease (PD), as well as cognitive disorders. As is well known, patients with PD show motor deficits such as slowed movement (bradykinesia), rigidity, and (resting) tremor, and they may also develop cognitive deficits such as dementia \cite{PD1,PD2,PD3,PD4}.

In this paper, we consider a spiking neural network of the BG, based on anatomical and physiological data obtained in rat-based works. 
Hence, we employ the rat-brain terminology. The BG take input from cortex through the input nuclei [striatum and subthalamic nucleus (STN)] and
make inhibitory output projection through the output nucleus [substantia nigra pars reticulata (SNr)] to the thalamus/brainstem \cite{Hump1,Man}.
Here, the principal input nucleus (striatum) takes cortical inputs from all over the cortex and also receives dopamine (DA), coming from
the substantia nigra pars compacta (SNc). Spine projection neurons (SPNs) are the only primary output neurons in the striatum, and they comprise up to 95 $\%$ of the whole striatal population, \cite{Str1,Str2}. Two kinds of SPNs with D1 and D2 receptors for the DA exist. Then, spiking activities of the two D1 and D2 SPNs are modulated differently by the DA \cite{SPN1,SPN2,CN6}.

Two competitive pathways, direct pathway (DP) and indirect pathway (IP), exist in the BG \cite{DIP1,DIP2,DIP3,DIP4}.
D1 SPNs in the striatum project inhibition directly to the output nucleus, SNr, via DP, and then the thalamus is disinhibited.
As a result, movement facilitation takes place. On the other hand, D2 SPNs are indirectly linked to the SNr via IP which crosses the intermediate
GP (globus pallidus) and the STN. In this case of IP, the firing activity of the SNr becomes intensified mainly due to excitation from the STN.
Consequently, spiking activity of the thalamus becomes reduced, leading to movement suppression.
In the case of normal DA level, DP is more active than IP, and an action is initiated (i.e., ``Go'' behavior occurs).
In contrast, for lower DA level, IP could be more active than DP, and then the action is withheld (i.e., ``No-Go'' behavior takes place).
In this way, DP and IP are also called the ``Go'' and ``No-Go'' pathways, respectively \cite{Frank1,Frank2,Go1,Go2}.

As is well known, a variety of functions of the BG may be done via ``balance'' between the ``Go'' DP and the ``No-Go'' IP, and such balance is regulated by the DA level \cite{Luo,Kandel,Squire,Bear}. Until now, diverse subjects for the BG have been studied in computational works. Various neuron models were used in the computational works;
(a) artificial neuron model of leaky-integrator type \cite{GPR1,GPR2,Hump3}, 
(b) point neuron function using the rate-coded output activation \cite{CN10,Frank1,Frank2},
(c) leaky integrate-and-fire model \cite{Hump1,Hump2}, 
(d) adaptive exponential integrate-and-fire model \cite{CN11,CN12,CN20},
(e)	oscillatory model for local field potentials \cite{CN18},
(f)	dendrite model \cite{CN8},
(g)	Hodgkin-Huxley type neuron model \cite{CN17,CNYu1,CNYu3,CNYu4}, and 
(h) Izhikevich neuron model \cite{SPN1,Str2,CN13,CN16,CN7,SPN2,Man,CN2,CN3,CN4,CN15,CN6,PD4,CN14,CN5,CN19,CN1,CN21,CNYu2}.

However, to the best of our knowledge, no quantitative analysis for balance between DP and IP was done.
[We also note that, using the terminology of balance could lead to misunderstanding; DP and IP are balanced (i.e., equal weighted in a strict sense?). To avoid such misunderstanding, we use the terminology of harmony.] To make clear the concept of such traditional ``balance,'' as a first time, we quantify competitive harmony (i.e., competition and cooperative interplay) between ``Go'' DP and ``No-Go'' IP. For such quantitative analysis, we first introduce the competition degree ${\cal C}_d$ between DP and IP. Here, ${\cal C}_d$ is provided by the ratio of strength of DP (${\cal S}_{DP}$) to strength of IP (${\cal S}_{IP}$); ${\cal C}_d = {\cal S}_{DP} / {\cal S}_{IP}$. The strengths of ``Go'' DP and ``No-Go'' IP, ${\cal S}_{DP}$ and ${\cal S}_{IP},$ are just the magnitudes of the total time-averaged presynaptic currents into the output nucleus, SNr, via ``Go'' DP and ``No-Go'' IP, respectively. Then, we can make quantitative analysis of harmony between DP and IP in terms of their competition degree ${\cal C}_d$, which could result in improving our understanding of the traditional ``balance'' so clearly and quantitatively.

We first consider the case of normal DA level of $\phi^*=0.3$.
For the tonic cortical input (3 Hz) in the resting state, a default state with ${\cal C}_d \simeq 1$ (i.e., DP and IP are nearly balanced in the strict sense) appears. In this default state with balanced DP and IP, the cells in the output nucleus, SNr, fire actively with the frequency 25.5 Hz, resulting in the locked state of the BG gate to the thalamus. Consequently, no movement occurs. On the other hand, for the phasic cortical input (10 Hz) in the phasically-active state, a healthy state with ${\cal C}_d^* = 2.82$ (i.e., DP is 2.82 times stronger than IP) is found to appear. In this healthy state, the firing frequency of the SNr becomes much reduced to 5.5 Hz from 25.5 Hz (default state), which leads to the opened state of the BG gate to the thalamus. Through this kind of competitive harmony between DP and IP, normal movement occurs in the healthy state, in contrast to the case of default state.

Next, we consider the case of reduced DA level, $\phi = \phi^*(=0.3)~x_{DA}$ ($1 > x_{DA} \geq 0$).
As $x_{DA}$ (i.e., fraction of the DA level) is decreased from 1, the competition degree ${\cal C}_d$ between DP and IP is found to decrease monotonically from ${\cal C}_d^*$, which leads to appearance of a pathological Parkinsonian state (i.e., PD) with reduced competition degree ${\cal C}_d$.
For the pathological Parkinsonian state, strength of IP (${\cal S}_{IP}$) is much increased than that for the normal healthy state, resulting in disharmony between DP and IP. Because of such break-up of harmony between DP and IP, arising from deficiency in DA production in the cells of the SNc \cite{PD5,PD6}, a pathological
Parkinsonian state with impaired movement occurs. Finally, we also investigate treatment of the Parkinsonian state through recovery of harmony between DP and IP.

This paper is organized as follows. In Sec.~\ref{sec:DGN}, we describe a spiking neural network for the BG. Then, in the main Sec.~\ref{sec:QA},
as a first time, we introduce the competition degree ${\cal C}_d$ between `Go'' DP and ``No-Go'' IP, and then quantify competitive harmony between the DP and the IP in terms of ${\cal C}_d$. Finally, a summary along with discussion on a possibility to get ${\cal C}_d$ experimentally is given in Sec.~\ref{sec:SUM}.

\section{Spiking Neural Network of The Basal Ganglia}
\label{sec:DGN}
We are concerned in spiking neural networks for the BG. In 2001, based on the functional anatomy proposed by Gurney et al. \cite{GPR1}, they developed an artificial neural network for the BG \cite{GPR2}. Later, in 2006, based on the anatomical and physiological data, Humphries et al.  \cite{Hump1} in the Gurney group developed a physiological neural model for the BG by employing the leaky integrate-and-fire neuron model with one dynamic variable \cite{LIF}. But, the effects of dopamine on the BG cells and synaptic currents were not considered there. In 2009, such effects of dopamine modulations on the striatal cells (D1 and D2 SPNs and fast-spiking interneurons) and the synaptic currents into the striatal cells were studied intensively by Humphries et al. \cite{Str2,SPN1} by using the Izhikevich neuron models \cite{Izhi1,Izhi2,Izhi3,Izhi4}. In 2017, Fountas and Shanahan \cite{CN6,CN7} extended the work of Humphries et al. \cite{Str2,SPN1} to the whole BG (including GP, STN, and SNr in addition to the striatal cells) by employing the Izhikevich neuron model, and studied oscillatory firing behaviors in the BG \cite{CN6,CN7} where dopamine effects were also considered. Also, in 2015 Mandali et al. \cite{Man} used the Izhikevich neuron models arranged on a 2D lattice for the BG cells and studied synchrony, exploration, and action selection. Recently, in 2021 Navarro-L\'{o}pez et al. \cite{CN1} also developed the BG-thalamo-cortical network (where the Izhikevich neuron models were also used), and investigated the BG-thalamo-cortical oscillatory activity. In some other spiking neural networks for the BG, instead of the Izhikevich neuron model, the adaptive exponential integrate-and-fire model with two dynamic variables \cite{AdEx} was used for the BG cells for study of signal enhancement by short-term plasticity \cite{CN11} and learning stimulus-action association \cite{CN20}.

In this section, based on the spiking neural networks (SNNs) for the BG developed in previous works \cite{SPN1,SPN2,CN6}, we make refinements on the BG SNN to become satisfactory for our study. This BG SNN is based on anatomical and physiological data of the BG as follows. For the framework of the BG SNN (e.g., number of BG cells and synaptic connection probabilities), refer to the anatomical works \cite{Ana1,Ana2,Ana3,Ana4}. For the intrinsic parameter values of single BG cells, we refer to
the physiological properties of the BG cells \cite{Phys1,Phys2,Phys3,Phys4,Phys5,Phys6,Phys7,Phys8,Phys9,Phys10,Phys11}. For the synaptic parameters (associated with synaptic currents), we also refer to the physiological works \cite{Phys12,Phys13,Phys14,Phys15,Phys16,Phys17,Phys18,Phys19,Phys20}.
Here, we consider the BG SNN, composed of D1/D2 SPNs, STN cells, GP cells, and SNr cells. 
For simplicity, within the striatum, only the dominant D1/D2 SPNs are taken into consideration (without considering a minor subpopulation of fast spiking interneurons), mutual interactions between the D1/D2 SPNs are not considered, and recurrent synaptic interactions are considered only in the case of GP. We employ the original Izhikevich neuron model with two dynamic variables for all the BG cells
\cite{Izhi1,Izhi2,Izhi3,Izhi4}. In addition, modulation effect of DA on D1/D2 SPNs and afferent synapses into the D1/D2 SPNs, the STN, and the GP is also taken into consideration \cite{SPN1,SPN2,CN6}. Here, we briefly present the governing equations for the population dynamics in the refined BG SNN; for details, refer to
Appendices \ref{app:A} and \ref{app:B}.

\begin{figure}
\includegraphics[width=0.85\columnwidth]{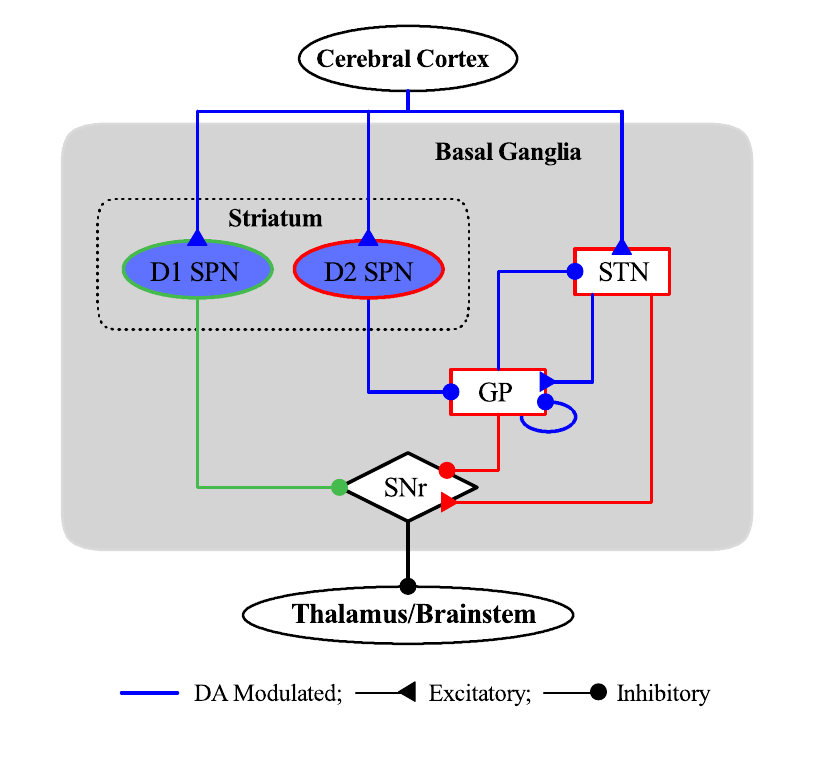}
\caption{Box diagram of a spiking neural network for the basal ganglia. We represent excitatory and inhibitory connections by lines with triangles and circles, respectively. We also denote dopamine-modulated cells and synaptic connections in blue color. There are two input nuclei to the BG, striatum and STN (subthalamic nucleus), taking the excitatory cortical input. Two types of inhibitory spine projection neurons (SPNs) exist in the striatum; D1 SPNs with the D1 receptors and
D2 SPNs with D2 receptors. The D1 SPNs project inhibition directly to the output nucleus SNr (substantia nigra pars reticulata) via DP (direct pathway; green color).
On the other hand, the D2 SPNs are indirectly linked to the SNr via IP (indirect pathway; red color) which crosses the GP (globus pallidus) and the STN.
Competition between DP and IP controls inhibitory output from the SNr to the thalamus/brainstem.
}
\label{fig:BGN}
\end{figure}

\subsection{Architecture of The Spiking Neural Network}
\label{subsec:SNN}
A box diagram of major cells and synaptic connections in the BG SNN is shown in Fig.~\ref{fig:BGN}.
This BG SNN is composed of the input nuclei (striatum and STN), the output nucleus (SNr), and the intermediate controller (GP).
Here, STN is the only excitatory cell in the BG, while all the other ones are inhibitory cells.
Particularly, we note that the SNr makes inhibitory output projections to the thalamus/brainstem, in contrast to the usual case of excitatory outputs.

Both striatum and STN take inputs from the cortex. Cortical inputs are modeled by employing 1,000 independent Poisson spike trains with spiking rate
$f_i$ $(i=1, \cdots, 1000)$. In the case of tonic cortical input in the resting state, $f=3$ Hz, while for the phasic cortical input in the phasically-active state, $f=10$ Hz, independently of $i$ \cite{Hump1,CI1,CI2,CI3,CI4,CI5,Str2,CN6,CN11}. Also, the principal input nucleus (striatum) receives the DA (coming from the SNc). Within the striatum, there are two types of SPNs with D1 and D2 receptors for the DA, comprising up to 95 $\%$ of the whole striatal population;
a minor subpopulation of fast spiking interneurons are not considered in the SNN \cite{Str1,Str2}. These D1 and D2 SPNs exhibit different spiking activities
due to DA modulation \cite{SPN1,SPN2,CN6}.

Two competitive pathways exist in the BG \cite{DIP1,DIP2,DIP3,DIP4,Frank1,Frank2,Go1,Go2}. The D1 SPNs make inhibitory projection to the output nucleus, SNr, directly through the ``Go'' DP (green color in Fig.~\ref{fig:BGN}). Then, the thalamus becomes disinhibited, leading to movement facilitation. In contrast, the D2 SPNs are indirectly linked to the SNr through the ``No-Go'' IP (red color in Fig.~\ref{fig:BGN}) which crosses the GP an the STN. Here, as an intermediate controller, the GP modulates the firing activity of the STN. In this case of IP, the firing activity of the SNr becomes enhanced mainly because of excitation from the STN. As a result, firing activity of the thalamus becomes reduced, resulting in movement suppression.
In this way, competition between ``Go'' DP (green) and ``No-Go'' IP (red) controls firing activity of the output nucleus, SNr.

\begin{table}
\caption{Numbers of BG cells, $N_{\rm X}$ [$X$ = D1 (SPN), D2 (SPN), STN, GP, and SNr] in the spiking neural network.}
\label{tab:NoBGCell}
\begin{tabular}{|c|c|}
\hline
\hspace{1cm} $N_{\rm D1}$ \hspace{1cm} & \hspace{1cm} 1,325  \hspace{1cm} \\
\hline
$N_{\rm D2}$ & 1,325 \\
\hline
$N_{\rm STN}$ & 14 \\
\hline
$N_{\rm GP}$ & 46 \\
\hline
$N_{\rm SNr}$ & 26 \\
\hline
\end{tabular}
\end{table}

\begin{table}
\caption{Synaptic connection probabilities $p_c^{(T,S)}$ from a presynaptic cell in the source population ($S$) to a postsynaptic cell in the target population ($T$).}
\label{tab:Pc}
\begin{tabular}{|c|c|}
\hline
& \hspace{1cm} $p_c^{(T,S)}$ \hspace{1cm} \\
\hline
D1 SPN $\rightarrow$ SNr & 0.033 \\
\hline
D2 SPN $\rightarrow$ GP & 0.033 \\
\hline
STN $\rightarrow$ GP & 0.3 \\
\hline
GP $\rightarrow$ GP & 0.1 \\
\hline
GP $\rightarrow$ STN & 0.1 \\
\hline
STN $\rightarrow$ SNr & 0.3 \\
\hline
GP $\rightarrow$ SNr & 0.1066 \\
\hline
\end{tabular}
\end{table}

Based on the anatomical information \cite{Ana3}, the numbers of the striatal cells, the STN cells, the SNr cells, and the GP cells in the BG are chosen.
Here, we consider a scaled-down SNN where the total number of striatal cells is $2,791$ which corresponds to $\frac {1}{1000}$ of the $2,791 \cdot 10^{3}$ striatal cells obtained in the BG of the rat. Thus, we make scaling down with ratio $10^{-3}$ for all the BG cells \cite{CN8,CN14}.
The total numbers of the BG cells are shown in Table~\ref{tab:NoBGCell}.
We note that 90-97 $\%$ of the whole striatal population is the major subpopulation of D1/D2 SPNs \cite{CN8};
here, we consider 95 $\%$. The remaining 5 $\%$ is a minor subpopulation of fast spiking interneurons (which are not considered in the SNN).

From the outside of the BG, the cortex (Ctx) makes external excitatory projections randomly to the D1/D2 SPNs and the STN cells; in this case, the
connection probabilities, $p_c^{(SPN,Ctx)}$ = 0.084 (8.4 $\%$) and $p_c^{({\rm STN,Ctx})}$ = 0.03 (3 $\%$), respectively \cite{CN6}.
As shown in Fig.~\ref{fig:BGN}, random synaptic connections between BG neurons are considered; we also consider random recurrent connections between GP cells.
Table \ref{tab:Pc} shows the synaptic connection probabilities $p_c^{(T,S)}$ from a presynaptic cell in the source population ($S$) to a postsynaptic cell in the target population ($T$) in the BG \cite{CN11}.

\subsection{Single Neuron Models, Synaptic Currents, and DA Effects}
\label{subsec:LIF-SC}
As single neuron models in the BG SNN, we use the Izhikevich spiking neuron model which is computationally efficient as well as biologically plausible \cite{Izhi1,Izhi2,Izhi3,Izhi4}, as in our previous works for spike-timing-dependent plasticity \cite{Kim1,Kim2,Kim3}.
The Izhikevich model matches neurodynamics through tuning its intrinsic parameters, instead of matching electrophysiological data, in contrast to
the Hodgkin-Huxley-type conductance-based models.

The BG SNN is composed of 5 populations of D1 SPNs, D2 SPNs, STN cells, GP cells, and SNr cells. The state of a cell in each population is characterized by its membrane potential $v$ and the slow recovery variable $u$ in the Izhikevich neuron model. Time-evolution of $v$ and $u$ is governed by three types of currents into the cell, $I_{ext}$ (external current), $I_{syn}$ (synaptic current), and $I_{stim}$ (stimulation current). Here, $I_{ext},$ $I_{syn}$, and
$I_{stim}$ represent stochastic external excitatory input from the external region (i.e., corresponding to the background part not considered in the modeling), the synaptic current, and the injected stimulation DC current, respectively. As the membrane potential reaches a threshold (i.e., spike cutoff value), firing a spike occurs, and then the membrane potential $v$  and the recovery variable $u$  are reset.

Detailed explanations on the Izhikevich neuron models for the D1/D2 SPNs, the STN cell, the GP cell, and the SNr cell
are presented in Appendix \ref{app:A} \cite{SPN1,SPN2,CN6}. Each Izhikevich neuron model has 9 intrinsic parameters which are
shown in Table \ref{tab:Single} in Appendix \ref{app:A}. These values are based on physiological properties of the BG cells  \cite{Phys1,Phys2,Phys3,Phys4,Phys5,Phys6,Phys7,Phys8,Phys9,Phys10,Phys11}.

Next, we consider the synaptic currents $I_{syn}$ into the BG cells. We follow the ``canonical'' formalism for $I_{syn}$
\cite{Cere1,Cere2,Hippo1,Hippo2,Hippo3,Hippo4}; for details, refer to Appendix \ref{app:B}.
There are two types of excitatory AMPA and NMDA receptor-mediated synaptic currents and one kind of inhibitory GABA receptor-mediated synaptic current.
Synaptic conductance for each synaptic current is provided by multiplication of maximum conductance per synapse, average number of afferent synapses, and fraction of open postsynaptic ion channels. We note that, postsynaptic ion channels are opened through binding of neurotransmitters to receptors in the target population.
A sum of the exponential-decay functions (controlled by the synaptic decay time constant and the synaptic latency time constant)
over presynaptic spikes provide temporal evolution of the fraction of open ion channels.
The synaptic parameter values (based on the physiological properties of the BG cells) for the maximum synaptic conductance, the synaptic decay time constant, the synaptic latency time constant, and the synaptic reversal potential for the synaptic currents are given in Table \ref{tab:SynParm} in Appendix \ref{app:B}
\cite{SPN2,CN6,CN11,Phys2,Phys12,Phys13,Phys14,Phys15,Phys16,Phys17,Phys18,Phys19,Phys20}.

Finally, we consider the DA effect on the BG SNN \cite{SPN1,SPN2,CN6}. Figure \ref{fig:BGN} shows influences of DA modulation on D1/D2 SPNs and synaptic currents into the D1/D2 SPNs, the STN cells, and the GP cells (blue color). The DA influences on the D1/D2 SPNs are well shown in the current-frequency (f-I) curves in Fig.~2A of Ref.~\cite{SPN1}. We note changes from the basic model (without DA; red) to the D1 (green) and the D2 (blue) SPN models.
Such changes occur due to different DA effects, depending on the D1 and D2 SPNs. D1 receptor activation has two opposing effects.
Due to a hyperpolarizing effect, activation threshold is increased in comparison to the bare case, while after threshold, the slope of the f-I curve increases rapidly because of another depolarizing effect. In contrast, in the case of D2 SPN, only the depolarizing effect occurs, leading to left-shift of the bare f-I curve. As a result of DA effects, excitatory cortical inputs into the D1 (D2) SPNs are upscaled (downscaled), as shown well in Fig.~2C of Ref.~\cite{SPN1}. All the other synaptic currents into the STN cells and the GP cells become downscaled due to DA effects. More details on the DA effects on the SPNs and synaptic currents are given in Appendices \ref{app:A} and \ref{app:B}, respectively.

\begin{figure}
\includegraphics[width=0.9\columnwidth]{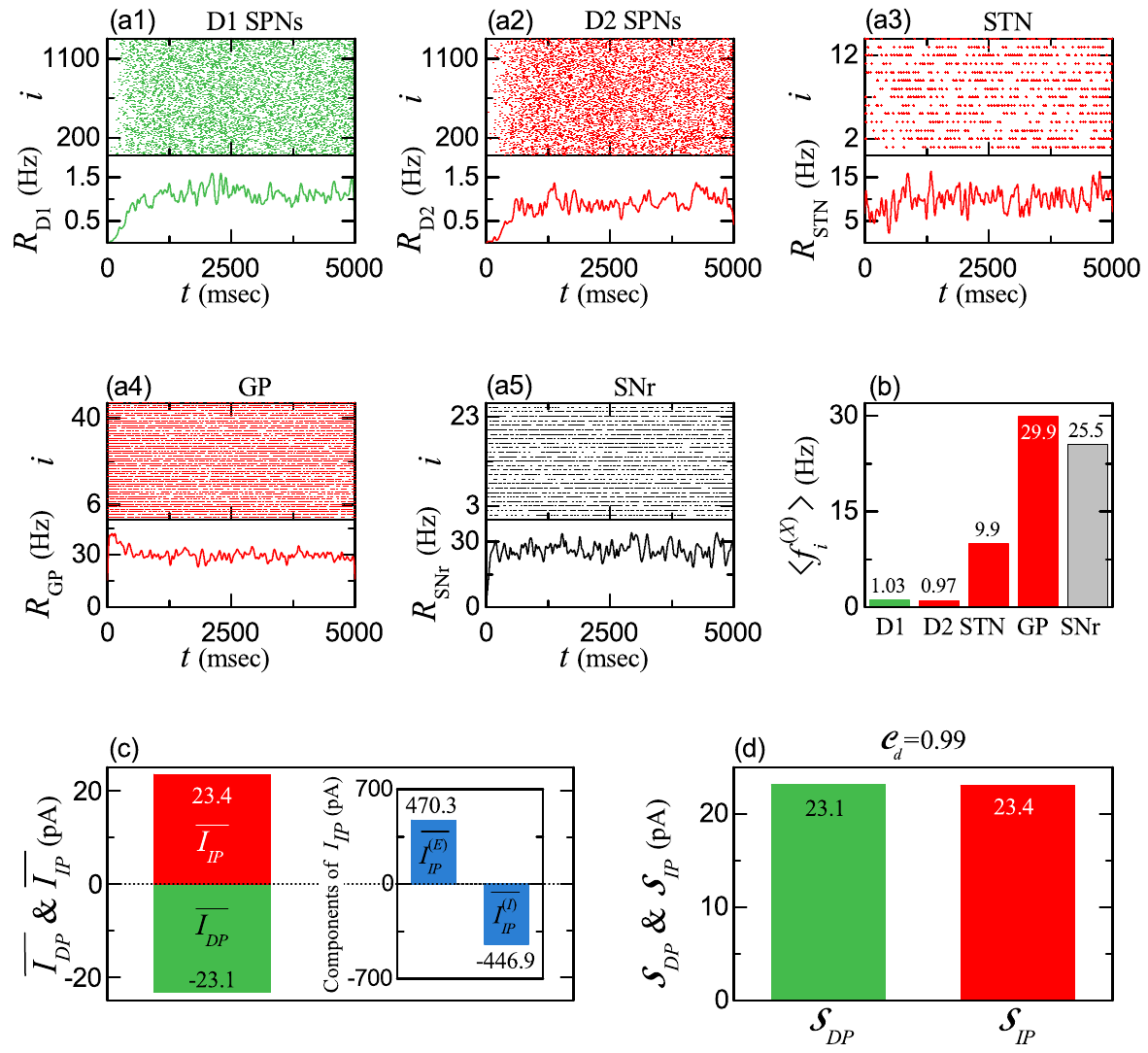}
\caption{Default basal ganglia state for the tonic cortical input (3 Hz) in the resting state and normal DA level $\phi = 0.3$.
Colors: parts, associated with DP (green), while parts, related to IP (red). Populations: $X=$ D1 (SPN), D2 (SPN), STN, GP, and SNr.
Raster plots of spikes and IPSRs (instantaneous population spike rates) $R_X (t)$ of (a1) D1 SPN, (a2) D2 SPN, (a3) STN, (a4) GP, and (a5) SNr cells. (b) Population-averaged mean firing rates (MFRs) $\langle f_i^{(X)} \rangle$ of D1 SPN, D2 SPN, STN, GP, and SNr) cells. (c) Time-averaged synaptic currents for DP ($\overline{I_{DP}}$) and IP ($\overline{I_{IP}}$). Inset shows the excitatory and the inhibitory components of the IP current, $\overline{I_{IP}^{(E)}}$ and $\overline{I_{IP}^{(I)}}$. (d) Strengths of DP (${\cal{S}}_{DP}$) and IP (${\cal{S}}_{IP}$). The competition degree ${\cal{C}}_d
 (={\cal{S}}_{DP}/{\cal{S}}_{IP})=0.99$.
}
\label{fig:Tonic}
\end{figure}

\section{Quantifying Competitive Harmony between DP and IP by Employing Their Competition Degree}
\label{sec:QA}
In this section, as a first time, we introduce the competition degree ${\cal C}_d$ between DP and IP. ${\cal C}_d$ is given by the ratio of strength of DP (${\cal S}_{DP}$) to strength of IP (${\cal S}_{IP}$) (i.e., ${\cal C}_d = {\cal S}_{DP} / {\cal S}_{IP}$). Then, we quantify competitive harmony (i.e., competition and cooperative interplay) between DP and IP by employing ${\cal C}_d$.

We first consider the normal DA level of $\phi=0.3$; $\phi_1$ (DA level for the D1 SPNs) = $\phi_2$ (DA level for the D2 SPNs) = $\phi$. For the tonic cortical input ($f=3$ Hz) in the resting state, a default state with ${\cal C}_d \simeq 1$ (i.e., DP and IP are nearly balanced) appears. In this default state, the BG gate to the thalamus is locked due to active firing activity of the cells in the output nucleus SNr, which results in no movement. On the other hand, for the phasic cortical input (10 Hz) in the phasically-active state, a healthy state with ${\cal C}_d^* = 2.82$ (i.e., DP is 2.82 times stronger than IP) appears. In this healthy state,
the BG gate to the thalamus becomes opened because the firing activity of the SNr cells is much reduced. Thus, normal movement occurs via competitive harmony
between DP and IP.

Next, we consider the case of decreased DA level, $\phi = \phi^*(=0.3)~x_{DA}$ ($1 > x_{DA} \geq 0$).
With reducing $x_{DA}$ from 1, the competition degree ${\cal C}_d$ between DP and IP decreases monotonically from ${\cal C}_d^*$ (= 2.82), which results in
appearance of a pathological state with reduced competition degree. In the pathological state, strength of IP (${\cal S}_{IP}$) is much increased than that for the normal healthy state, leading to disharmony between DP and IP. Due to break-up of harmony between DP and IP, arising from deficiency in DA production in the cells of the SNc \cite{PD5,PD6}, PD with impaired movement occurs. Finally, we also study treatment of the pathological state with PD via recovery of harmony between DP and IP.

\subsection{Healthy States with Harmony between DP and IP}
\label{subsec:HS}
We consider the case of normal DA level of $\phi=0.3$ for the D1 and D2 SPNs.
As explained in Sec.~\ref{subsec:SNN}, cortical inputs are modeled in terms of 1,000 independent Poisson spike trains with firing rate $f$.
We first consider the case of tonic cortical input with $f=3$ Hz in the resting state, observed in experimental works \cite{CI1,CI2,CI3,CI4,CI5}
and used in computational works \cite{Hump1,Str2,CN6,CN11}. 
Population firing activity of BG cells may be well visualized in the raster plot of spikes which is a collection of spike trains of individual BG cells.
Figures \ref{fig:Tonic}(a1)-\ref{fig:Tonic}(a5) show the raster plots of spikes for D1 SPNs (green), D2 SPNs (red), STN cells (red), GP cells (red), and SNr cells, respectively; color of D1 SPNs, associated with DP is green, while color of BG cells related to IP is red.

As a collective quantity exhibiting population behaviors, we use an IPSR (instantaneous population spike rate) which could be obtained from the raster plot of spikes \cite{IPSR1,IPSR2,IPSR3,IPSR4,IPSR5}. In this case, each spike in the raster plot is convoluted with a kernel function $K_h(t)$ to obtain a smooth estimate of IPSR $R_X(t)$ in the $X$ population ($X=$ D1 (SPN), D2 (SPN), STN, GP, and SNr) \cite{Kernel}:
\begin{equation}
R_X(t) = \frac{1}{N_X} \sum_{i=1}^{N_X} \sum_{s=1}^{n_i^{(X)}} K_h (t-t_{s,i}^{(X)}).
\label{eq:IPSR}
\end{equation}
Here, $N_X$ is the number of the cellss, and $n_i^{(X)}$ and $t_{s,i}^{(X)}$ are the total number of spikes and the $s$th spiking time of the $i$th cell,
respectively. As the kernel function, we employ a Gaussian function of band width $h$:
\begin{equation}
K_h (t) = \frac{1}{\sqrt{2\pi}h} e^{-t^2 / 2h^2}, ~~~~ -\infty < t < \infty,
\label{eq:Gaussian}
\end{equation}
where the band width $h$ of $K_h(t)$ is 20 msec. The IPSRs $R_X(t)$ for $X=$ D1 (SPN), D2 (SPN), STN, GP, and SNr are also shown in
Figs.~\ref{fig:Tonic}(a1)-\ref{fig:Tonic}(a5), respectively.

As shown in Fig.~\ref{fig:Tonic}(b), population-averaged mean firing rates (MFRs) of BG cells, $\langle f_i ^{(X)}\rangle$, for the tonic case are 1.03, 0.97, 9.9, 29.9, and 25.5 Hz for $X=$ D1 (SPN), D2 (SPN), STN, GP, and SNr, respectively \cite{Hump1,CN6,CN11}; $f_i^{(X)}$ is the MFR of the $i$th cell in the $X$ population and $\langle \cdots \rangle$ denotes the population average over all cells. For details, refer to Table \ref{tab:Spon} in Appendix \ref{app:A}.
In this case of default BG state, the D1 and D2 SPNs in the input nucleus (striatum) are nearly silent.
On the other hand, the output SNr cells fire very actively, and hence the BG gate to the thalamus becomes locked, leading to no movement.

There are two types of synaptic currents into the (output) SNr cells, $I_{DP}$ and $I_{IP}$, via DP (green) and IP (red) in Fig.~\ref{fig:BGN}, respectively.
For details of synaptic  currents, refer to Appendix \ref{app:B}; refer to Eq.~(\ref{eq:I}) for all the currents into the cell.
Here, the DP current, $I_{DP}(t),$ is just the (inhibitory) synaptic current from the D1 SPNs to the SNr cells:
\begin{equation}
  I_{DP}(t) = - I_{syn}^{({\rm SNr,D1})}(t).
\label{eq:DPC}
\end{equation}
The IP current, $I_{IP}(t),$ consists of the excitatory component, $I_{IP}^{(E)}(t),$ and the inhibitory component, $I_{IP}^{(I)}(t):$
\begin{equation}
  I_{IP}(t) = I_{IP}^{(E)}(t) + I_{IP}^{(I)}(t).
\label{eq:IPC}
\end{equation}
Here, $I_{IP}^{(E)}(t)$ [$I_{IP}^{(I)}(t)$] is just the synaptic current from the STN (GP) to the SNr:
\begin{equation}
  I_{IP}^{(E)}(t) = - I_{syn}^{({\rm SNr,STN})}(t)~~~{\rm and}~~~  I_{IP}^{(I)}(t) = - I_{syn}^{({\rm SNr,GP})}(t).
\label{IPCEI}
\end{equation}

We note that, firing activity of the (output) SNr cells is determined via competition between DP current [$I_{DP}(t)$] and IP current
[$I_{IP}(t)$] into the SNr. The strengths of DP and IP, ${\cal S}_{DP}$ and ${\cal S}_{IP}$, are given by the magnitudes of their respective time-averaged synaptic currents:
\begin{equation}
  {\cal S}_{DP} = |\overline{I_{DP}(t)}|~~~{\rm and}~~~{\cal S}_{IP} = |\overline{I_{IP}(t)}|,
\label{eq:Strength}
\end{equation}
where the overline represents the time averaging and $| \cdots |$ denotes the absolute magnitude.
Then, we introduce the competition degree ${\cal C}_d$ between DP and IP, given by the ratio of ${\cal S}_{DP}$ to ${\cal S}_{IP}$:
\begin{equation}
{\cal C}_d =  \frac {{\cal S}_{DP}} {{\cal S}_{IP}}.
\label{eq:CD}
\end{equation}
For ${\cal C}_d=1$, DP and IP are balanced, and the SNr cells fire actively with the MFR 25.5 Hz. Hence, the thalamic cells become silent,
leading to no movement. In the case of ${\cal C}_d >1$, DP is more active than IP, and hence, the firing activities of SNr cells are suppressed than the balanced state with ${\cal C}_d=1$. Thus, the BG gate to the thalamus becomes open, leading to movement facilitation. On the other hand, for ${\cal C}_d <1,$ IP is more active than DP, and hence, the firing activity of SNr cells are enhanced than the balanced state with ${\cal C}_d=1$. Thus, the BG gate to the thalamus becomes locked, resulting in movement suppression.

\begin{figure}
\includegraphics[width=0.85\columnwidth]{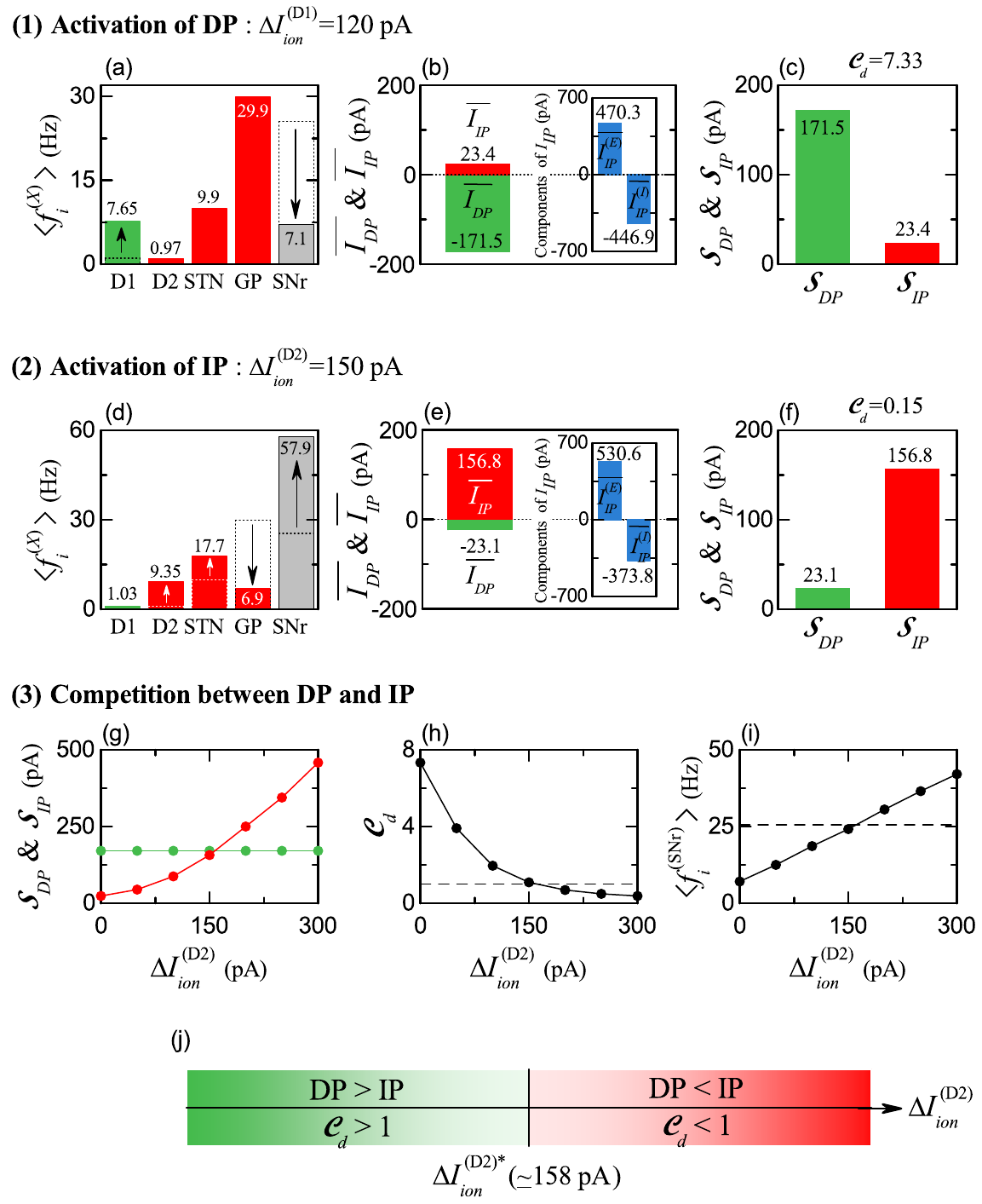}
\caption{Activations of DP and IP. Colors: parts, associated with DP (green), while parts, related to IP (red). Populations: $X=$ D1 (SPN), D2 (SPN), STN, GP, and SNr. (1) Activation of DP for $\Delta I_{ion}^{({\rm D1})}=$ 120 pA: (a) Population-averaged MFRs $\langle f_i^{(X)} \rangle$
of D1 SPN, D2 SPN, STN, GP, and SNr cells. Dotted boxes for D1 SPN and SNr represent population-averaged MFRs for $\Delta I_{ion}^{({\rm D1})}=$ 0 pA, respectively. (b) Time-averaged synaptic current for DP $\overline{I_{DP}}$ and IP $\overline{I_{IP}}$. Inset shows the excitatory and the inhibitory components of the IP current, $\overline{I_{IP}^{(E)}}$ and $\overline{I_{IP}^{(I)}}$. (c) Strengths of DP (${\cal{S}}_{DP}$) and IP (${\cal{S}}_{IP}$). The competition degree ${\cal{C}}_d = 7.33$.
(2) Activation of IP for $\Delta I_{ion}^{({\rm D2})}=$ 150 pA: (d) Population-averaged MFR $\langle f_i^{(X)} \rangle$
of D1 SPN, D2 SPN, STN, GP, and SNr cells. Dotted boxes for D2 SPN, STN, GP, and SNr represent population-averaged MFRs for $\Delta I_{ion}^{({\rm D2})}=$ 0 pA, respectively. (e) Time-averaged synaptic current for DP $\overline{I_{DP}}$ and IP $\overline{I_{IP}}$. Inset shows the excitatory and the inhibitory components of the IP current, $\overline{I_{IP}^{(E)}}$ and $\overline{I_{IP}^{(I)}}$. (f) Strengths of DP (${\cal{S}}_{DP}$) and IP (${\cal{S}}_{IP}$). The competition degree ${\cal{C}}_d = 0.15$.
(3) Competition between DP and IP for $\Delta I_{ion}^{({\rm D1})}=$ 120 pA: (g) Plots of strengths of DP (${\cal{S}}_{DP}$) and IP (${\cal{S}}_{IP}$) versus $\Delta I_{ion}^{({\rm D2})}$. (h) Plot of the competition degree ${\cal{C}}_d$ versus $\Delta I_{ion}^{({\rm D2})}$. Horizontal dashed line represents ${\cal{C}}_d=1$. (i) Plot of population-averaged MFR of SNr $\langle f_i^{({\rm SNr})} \rangle$ versus $\Delta I_{ion}^{({\rm D2})}$. Horizontal dashed line represents $\langle f_i^{({\rm SNr})} \rangle =$ 25.5 Hz for $\Delta I_{ion}^{({\rm D1})}=\Delta I_{ion}^{({\rm D2})}=$ 0 pA. (j) Bar diagram for the competition between DP and IP. Green and red represent ${\rm DP} > {\rm IP}$ and ${\rm IP} > {\rm DP}$, respectively.
}
\label{fig:Acti}
\end{figure}

Hereafter, we employ the above competition degree ${\cal C}_d$ between DP and IP and make quantitative analysis for all the default, healthy, and pathological states occurring in the BG.
Figure \ref{fig:Tonic}(c) shows the time-averaged DP (green) and IP (red) currents for the tonic cortical input, $\overline {I_{DP}(t)}=-23.1$ and
$\overline {I_{IP}(t)}=23.4;$ in the case of IP current, time-averaged values (blue) of their excitatory and inhibitory components are also given,
$\overline { I_{IP}^{(E)}(t)} = 470.3$ and $\overline { I_{IP}^{(I)}(t)} = -446.9$.  Thus, the strengths of DP and IP become ${\cal S}_{DP}=23.1$ and
${\cal S}_{IP}=23.4,$ respectively, as shown in Fig.~\ref{fig:Tonic}(d). Consequently, the competition degree between DP and IP is ${\cal C}_d=0.99$ (i.e., DP and IP are nearly balanced). In this way, a default state with ${\cal C}_d \simeq 1$ appears for the tonic cortical input.
In this case, the (output) SNr cells fire very actively at $\langle f_i^{({\rm SNr})} \rangle =25.5$ Hz and make strong inhibitory projections to the thalamic cells. Thus, the BG gate to the thalamus is locked for the tonic cortical input, resulting in no movement.

We are also concerned about activation and deactivation of cells in the target population $X$ \cite{OG1,OG2} which could be used for treatment of pathological states. Optogenetics is a technique that combines optics and genetics to control the activity of target cells in living organisms, typically using light-sensitive proteins called opsins. The target cells are genetically modified to express these opsins (i.e., fusion of the opsins into the target cells). When the opsins are activated by specific wavelengths of light, variation in the intrinsic ionic currents of the cells in the target population $X$, $\Delta I_{ion}^{(X)}$, occurs.
When $\Delta I_{ion}^{(X)}$ is positive (negative), firing activity of the target cells is increased (decreased), leading to their activation (deactivation).

The governing equations for evolution of dynamical states of individual Izhikevich neurons in the $X$ population are given in Eqs.~(\ref{eq:GE1}) and (\ref{eq:GE2}) in Appendix \ref{app:A}. Time evolutions of the dynamical variables are governed by the current $I_i^{(X)}(t)$ of Eq.~(\ref{eq:I}) in Appendix \ref{app:A} into the $i$th cell in the $X$ population. Here, to simulate the effect of optogenetics, in addition to the current $I_i^{(X)}(t)$, we add variation of the intrinsic ionic currents of the target cells via the light stimulation, $\Delta I_{ion}^{(X)}(t)$ in Eq.~(\ref{eq:GE1}).

Light stimulation for optogenetics is applied on target cells in the case of tonic cortical input (3 Hz).
As target cells, we first consider D1 SPNs. With increasing the intensity of light stimulation, magnitude of $\Delta I_{ion}^{({\rm D1})}$ increases.
As an example, Figs.~\ref{fig:Acti}(a)-\ref{fig:Acti}(c) show the effects of optogenetics for $\Delta I_{ion}^{({\rm D1})} = 120$ pA.
The MFR $\langle f_i^{({\rm D1})} \rangle$ of D1 SPNs, associated with DP, is much increased to 7.65 Hz from 1.03 Hz (default state);
MFRs of other cells (D2 SPNs, STN, GP), related to IP, remain unchanged (i.e., same as those for the default state) [Fig.~\ref{fig:Acti}(a)].
Thus, DP becomes activated via activation of D1 SPNs. Then, firing activities of the output SNr cells are much suppressed; the MFR of SNr cells, $\langle f_i^{({\rm SNr})} \rangle,$ is much reduced from 25.5 Hz (default state) to 7.1 Hz (down-arrow).
In this case, strength of the DP, ${\cal S}_{DP}$ is much increased to 171.5 from 23.1 (default state) [Figs.~\ref{fig:Acti}(b) and \ref{fig:Acti}(c)].
Thus, the competition degree ${\cal C}_d$ between DP and IP becomes 7.33 which is much larger than that (= 0.99) for the default state.
Consequently, through activation of DP, the BG gate to thalamus becomes opened, leading to movement facilitation.

\begin{figure}
\includegraphics[width=0.9\columnwidth]{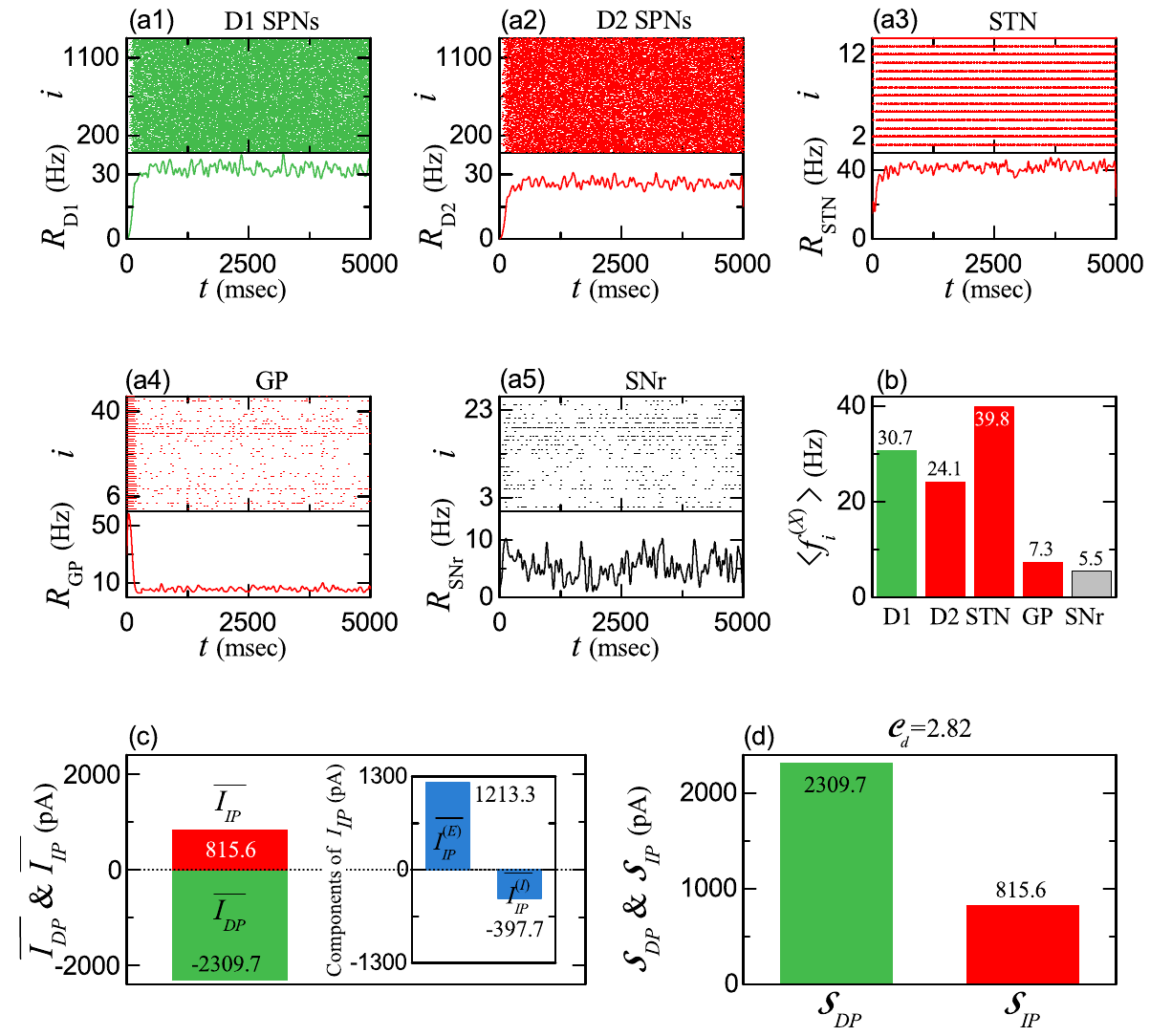}
\caption{Healthy state for the phasic cortical input (10 Hz) in the phasically-active state and normal DA level $\phi=$ 0.3.
 Colors: parts, associated with DP (green), while parts, related to IP (red). Populations: $X=$ D1 (SPN), D2 (SPN), STN, GP, and SNr.
 Raster plots of spikes and IPSRs $R_X (t)$ of (a1) D1 SPN, (a2) D2 SPN, (a3) STN, (a4) GP, and (a5) SNr cells. (b) Population-averaged MFR of D1 SPN, D2 SPN, STN, GP, and SNr cells. (c) Time-averaged synaptic current for DP ($\overline{I_{DP}}$)  and IP ($\overline{I_{IP}}$). Inset shows the excitatory and the inhibitory components of the IP current, $\overline{I_{IP}^{(E)}}$ and $\overline{I_{IP}^{(I)}}$. (d) Strengths of DP (${\cal{S}}_{DP}$) and IP (${\cal{S}}_{IP}$). The competition degree ${\cal{C}}_d^* (={\cal{S}}_{DP}/{\cal{S}}_{IP})=2.82$.
}
\label{fig:Phasic}
\end{figure}

Next, D2 SPNs are considered as target cells for optogenetics.
As an example, Figs.~\ref{fig:Acti}(d)-\ref{fig:Acti}(f) show the effects of optogenetics for $\Delta I_{ion}^{({\rm D2})} = 150$ pA.
The MFRs $\langle f_i^{(X)}\rangle$ of the cells [$X=$ D2 (SPN), GP, STN], associated with IP, are changed, while the MFR
$\langle f_i^{({\rm D1})} \rangle$ of D1 SPNs, related to DP, remains unchanged [Fig.~\ref{fig:Acti}(d)]. $\langle f_i^{({\rm D2})}\rangle$ of D2 SPNs is increased to 9.35 Hz from 0.97 Hz (default state). Due to increased inhibitory projections from D2 SPNs, $\langle f_i^{({\rm GP})}\rangle$ of GP cells is decreased to 6.9 Hz from 29.9 Hz (default state). Because of reduced firing activity of GP cells,
$\langle f_i^{({\rm STN})}\rangle$ of the STN cells increases to 17.7 Hz from 9.9 Hz (default state). Thus, the strength of IP, ${\cal S}_{IP}$, becomes much increased to
156.8 from 23.4 (default state)[Figs.~\ref{fig:Acti}(e) and \ref{fig:Acti}(f)].
In this way, IP is activated. Then, the competition degree, ${\cal C}_d$, between DP and IP becomes 0.15 which is much smaller than
that (= 0.99) for the default state. As a result, via activation of IP, the BG gate to thalamus is locked, resulting in movement suppression.

\begin{figure*}
\includegraphics[width=1.7\columnwidth]{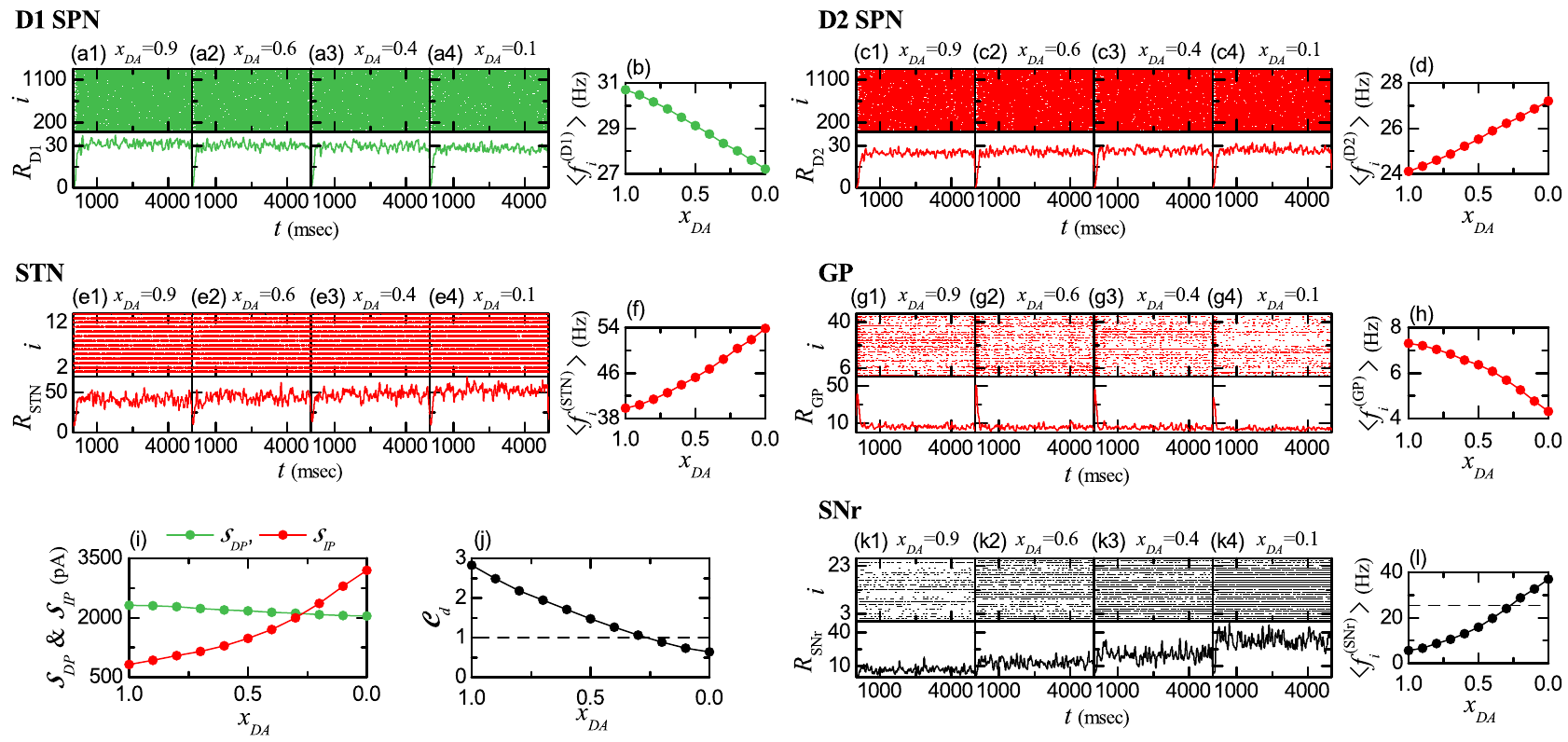}
\caption{Pathological state for the phasic cortical input (10 Hz) in the phasically-active state.
Colors: parts, associated with DP (green), while parts, related to IP (red).
(a1)-(a4) Raster plots of spikes and IPSRs $R_{\rm D1} (t)$ of D1 SPNs when $x_{DA}$ (fraction of DA level) is 0.9, 0.6, 0.4, and 0.1, respectively. (b) Population-averaged MFR $\langle f_i^{({\rm D1})} \rangle$ of D1 SPNs versus $x_{DA}$. (c1)-(c4) Raster plots of spikes and IPSRs $R_{\rm D2} (t)$ of D2 SPNs when $x_{DA}$  is 0.9, 0.6, 0.4, and 0.1, respectively. (d) Population-averaged MFR $\langle f_i^{({\rm D2})} \rangle$ of D2 SPNs versus $x_{DA}$. (e1)-(e4) Raster plots of spikes and IPSRs $R_{\rm STN} (t)$ of STN cells when $x_{DA}$ is 0.9, 0.6, 0.4, and 0.1, respectively. (f) Population-averaged MFR $\langle f_i^{({\rm STN})} \rangle$ of STN cells versus $x_{DA}$. (g1)-(g4) Raster plots of spikes and IPSRs $R_{\rm GP} (t)$ of GP cells when $x_{DA}$ is 0.9, 0.6, 0.4, and 0.1, respectively. (h) Population-averaged MFR $\langle f_i^{({\rm GP})} \rangle$ of GP cells versus $x_{DA}$. (i) Plots of strengths of DP (${\cal{S}}_{DP}$) and IP (${\cal{S}}_{IP}$) versus $x_{DA}$. (j) Plot of the competition degree ${\cal{C}}_d (={\cal{S}}_{DP}/{\cal{S}}_{IP})$ versus $x_{DA}$. Horizontal dashed line represents ${\cal{C}}_d =1$. (k1)-(k4) Raster plots of spikes and IPSRs $R_{\rm SNr} (t)$ of SNr cells when $x_{DA}$ is 0.9, 0.6, 0.4, and 0.1, respectively.  (l) Population-averaged MFR $\langle f_i^{({\rm SNr})} \rangle$ of SNr cells versus $x_{DA}$. Horizontal dashed line represents $\langle f_i^{({\rm SNr})} \rangle =$ 25.5 Hz for the default tonic state.
}
\label{fig:PD}
\end{figure*}

As a 3rd case, we study competition between DP and IP via light stimulation on both  D1 and D2 SPNs.
For simplicity, activation of D1 SPNs is fixed for $\Delta I_{ion}^{({\rm D1})} = 120$ pA; in this case, strength of the DP, ${\cal S}_{DP}$ is 171.5.
By increasing $\Delta I_{ion}^{({\rm D2})}$ from 0, competition between DP and IP is investigated.
Figures \ref{fig:Acti}(g)-\ref{fig:Acti}(i) show well the effects of optogenetics on their competition.
As $\Delta I_{ion}^{({\rm D2})}$ is increased from 0, the strength of IP, ${\cal S}_{IP}$ is found to monotonically increase from 23.4 [Fig.~\ref{fig:Acti}(g)].
Due to monotonic increase in ${\cal S}_{IP}$, the competition degree ${\cal C}_d$ between DP and IP decreases monotonically from 7.33 [Fig.~\ref{fig:Acti}(h)],
and the MFR of the (output) SNr cells, $\langle f_i^{({\rm SNr})} \rangle$, increases monotonically from 7.1 Hz [Fig.~\ref{fig:Acti}(i)].
We note that, when passing a threshold, $\Delta I_{ion}^{({\rm D2}*)} (\simeq$ 158 pA), ${\cal S}_{IP}$ becomes the same as ${\cal S}_{DP}$.
Figure~\ref{fig:Acti}(j) shows a diagram for competition between DP and IP.
For $\Delta I_{ion}^{({\rm D2})} < \Delta I_{ion}^{({\rm D2}*)}$, ${\cal S}_{DP}$ of DP is larger than  ${\cal S}_{IP}$ of IP (i.e., ${\cal C}_d >1$),
and then the MFR of SNr cells, $\langle f_i^{({\rm SNr})} \rangle$, becomes smaller than that (= 25.5 Hz) for the default state.
Consequently, the BG gate to thalamus is opened, leading to movement facilitation.
On the other hand, for $\Delta I_{ion}^{({\rm D2})} > \Delta I_{ion}^{({\rm D2}*)}$, ${\cal S}_{IP}$ of IP is larger than ${\cal S}_{DP}$ of DP,
and then the mean firing rate of SNr cells, $\langle f_i^{({\rm SNr})} \rangle$, becomes larger than that (= 25.5 Hz) for the default state.
As a result, the BG gate to thalamus is locked, resulting to movement suppression.

From now on, we consider the case of phasic cortical input with $f=10$ Hz in the phasically-active state, in contrast to the above case of tonic cortical input
with $f=$ 3 Hz in the resting default state \cite{CI1,CI2,CI3,CI4,CI5,Str2,CN6,CN11}.
Population firing behaviors of the BG cells may be well seen in the raster plots of spikes and they may also be characterized well in terms of their IPSRs.
Figures \ref{fig:Phasic}(a1)-\ref{fig:Phasic}(a5) show the raster plots of spikes and the IPSRs $R_X(t)$ for $X=$ D1 SPN (green), D2 SPN (red), STN (red), GP  (red), and SNr, respectively.

As shown in Fig.~\ref{fig:Phasic}(b), population-averaged MFRs of BG cells, $\langle f_i ^{(X)}\rangle$, for the phasic case are 30.7, 24.1, 39.8, 7.3, and 5.5 Hz for $X=$ D1 (SPN), D2 (SPN), STN, GP, and SNr, respectively.
We note that $\langle f_i ^{({\rm D1})}\rangle$ and $\langle f_i ^{({\rm D2})}\rangle$ of D1 and D2 SPNs are much larger than those for the tonic default case
with $\langle f_i^{({\rm D1})} \rangle =1.03$ Hz and $\langle f_i^{({\rm D2})} \rangle=0.97$ Hz.
As a result of activation of both D1 SPNs and D2 SPNs, both DP and IP become activated.
In the case of IP, $\langle f_i^{({\rm GP})} \rangle$ of GP cells is reduced from that (= 29.9 Hz) for the resting default state due to strong inhibition from the D2 SPNs, and $\langle f_i^{({\rm STN})} \rangle$ of STN cells is increased from that (= 9.9 Hz) for the default state because of reduced inhibition from the GP cells.
Through competition between DP and IP, the firing activities of the output SNr cells are suppressed [i.e. their
MFR, $\langle f_i^{({\rm SNr})} \rangle$, is reduced to 5.5 Hz from 25.5 Hz (default state)]. Due to reduced activity of SNr cells, the thalamus becomes disinhibited. Thus, the BG gate to the thalamus is opened, leading to movement facilitation.

We make quantitative analysis of DP and IP currents, $I_{DP}$ and $I_{IP}$, into the SNr. The strengths of DP and IP, ${\cal S}_{DP}$ and ${\cal S}_{IP}$, given by the magnitudes of time-averaged DP current ($I_{DP}$) and IP current ($I_{IP}$), are 2309.7 and 815.6, respectively [Figs.~\ref{fig:Phasic}(c) and \ref{fig:Phasic}(d)]. They are much increased from ${\cal S}_{DP}$ (= 23.1) and ${\cal S}_{IP}$ (= 23.4) in the default state. But, we note that, in the case of phasic cortical input (10 Hz), ${\cal S}_{DP}$ is much more increased than ${\cal S}_{IP}$.
Hence, the competition degree ${\cal C}_d^*$ between DP and IP, given by the ratio of ${\cal S}_{DP}$ to ${\cal S}_{IP},$ becomes 2.82
(i.e., DP is 2.82 times stronger than IP), in contrast to the default state with ${\cal C}_d \simeq 1$ (i.e., DP and IP are nearly balanced).
As a result of more activeness of DP, the MFR of the output SNr cells, $\langle f_i^{({\rm SNr})} \rangle$, becomes much decreased to 5.5 Hz from 25.5 Hz (default state). Consequently, in this healthy state with ${\cal C}_d^* = 2.82$, the BG gate to the thalamus becomes opened, leading to facilitation of normal movement, via competitive harmony (i.e., competition and cooperative interplay) between DP and IP.

\begin{figure*}
\includegraphics[width=1.3\columnwidth]{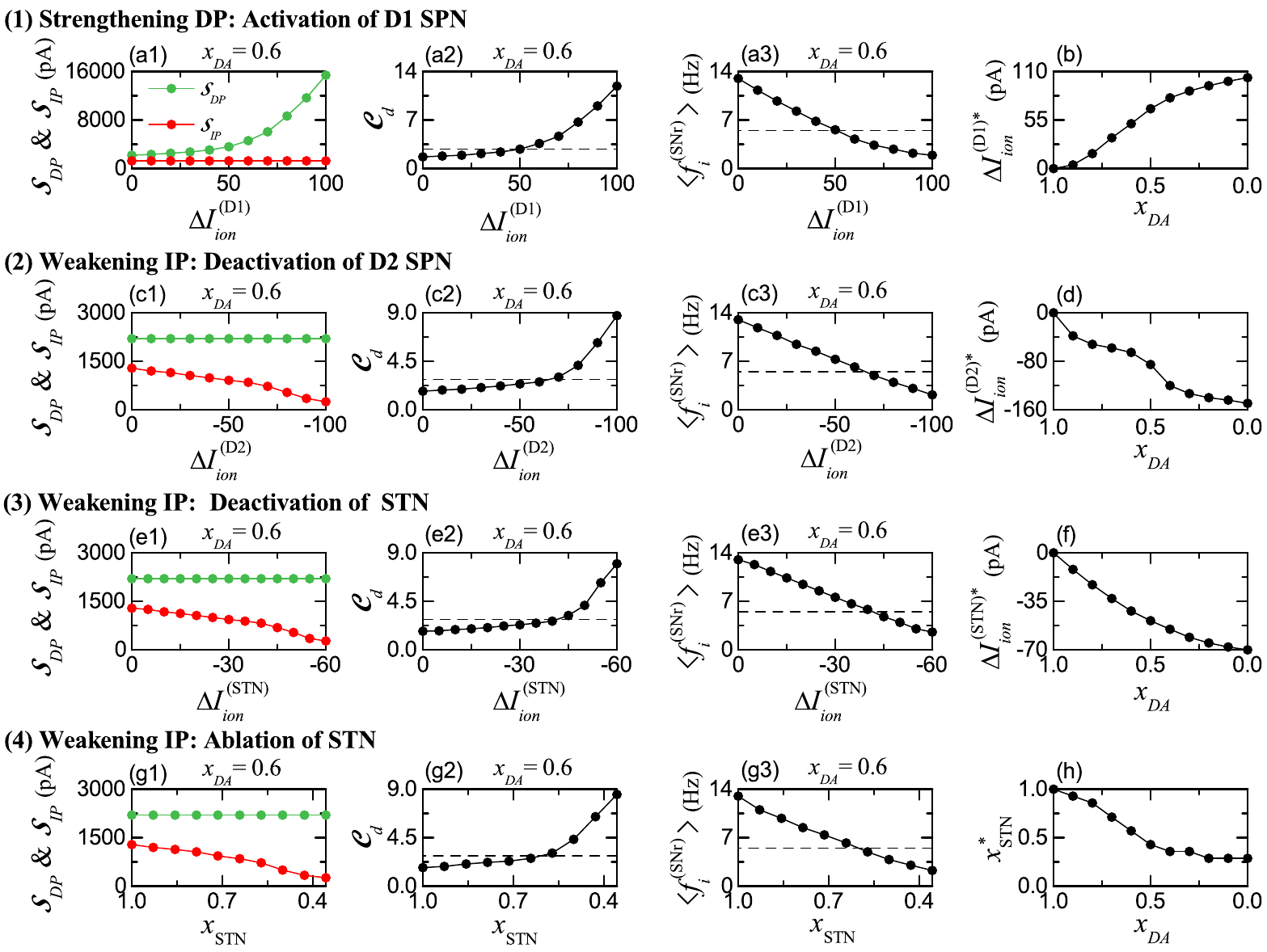}
\caption{Treatment of pathological states. Colors: parts, associated with DP (green), while parts, related to IP (red).
(1) Strengthening DP by activation of D1 SPN. Plots of (a1) ${\cal{S}}_{DP}$ (strength of DP) and ${\cal{S}}_{IP}$ (strength of IP), (a2) ${\cal{C}}_d$
(competition degree), and (a3) $\langle f_i^{({\rm SNr})} \rangle$ (MFR of SNr cells) versus $\Delta I_{ion}^{({\rm D1})}$ for $x_{DA}=0.6$. (b) Plot of
$\Delta I_{ion}^{({\rm D1})*}$ (threshold) versus $x_{DA}$.
(2) Weakening IP by deactivation of D2 SPN. Plots of (c1) ${\cal{S}}_{DP}$ and ${\cal{S}}_{IP}$, (c2) ${\cal{C}}_d$, and (c3) $\langle f_i^{({\rm SNr})} \rangle$ versus $\Delta I_{ion}^{({\rm D2})}$ for $x_{DA}=0.6$. (d) Plot of $\Delta I_{ion}^{({\rm D2})*}$ (threshold) versus $x_{DA}$.
(3) Weakening IP by deactivation of STN. Plots of (e1) ${\cal{S}}_{DP}$ and ${\cal{S}}_{IP}$, (e2) ${\cal{C}}_d$, and (e3) $\langle f_i^{({\rm SNr})} \rangle$ versus $\Delta I_{ion}^{({\rm STN})}$ for $x_{DA}=0.6$. (f) Plot of $\Delta I_{ion}^{({\rm STN})*}$ (threshold) versus $x_{DA}$.
(4) Weakening IP by ablation of STN cells. Plots of (g1) ${\cal{S}}_{DP}$ and ${\cal{S}}_{IP}$, (g2) ${\cal{C}}_d$, and (g3) $\langle f_i^{({\rm SNr})} \rangle$ versus $x_{\rm STN}$ for $x_{DA}=0.6$. (h) Plot of $x_{\rm STN}^*$ (threshold) versus $x_{DA}$. Horizontal dashed lines in (a2), (c2), (e2), and (g2) represent ${\cal{C}}_d^*$ (= 2.82) for the healthy state when $x_{DA}=1$. Horizontal dashed lines in (a3), (c3), (e3), and (g3) represent $\langle f_i^{({\rm SNr})} \rangle$ (= 5.5 Hz) for the healthy state when $x_{DA}=1$.
}
\label{fig:Treat}
\end{figure*}

\subsection{Pathological Parkinsonian States with Disharmony between DP and IP}
\label{subsec:PS}
In this subsection, we consider the case of reduced DA level, $\phi = \phi^*(=0.3)~x_{DA}$ ($1 > x_{DA} \geq 0$); $\phi^*$ (=0.3) is the normal DA level
\cite{PD5,PD6}. With decreasing the fraction of DA level, $x_{DA}$, we make quantitative analysis of strengths of DP (${\cal S}_{DP}$) and IP (${\cal S}_{IP}$), their competition degree ${\cal C}_d$, and (population-averaged) MFRs, $\langle f_i^{(X)} \rangle$ of the BG cells in the $X$ populations [$X=$ D1 (SPN), D2 (SPN), STN, GP, and SNr], in the case of phasic cortical input with $f=10$ Hz in the phasically-active state.

For D1 SPNs, raster plots of spikes and IPSRs are shown in Figs.~\ref{fig:PD}(a1)-\ref{fig:PD}(a4) for $x_{DA}$= 0.9, 0.6, 0.4, and 0.1, respectively.
Their (population-averaged) MFR $\langle f_i^{({\rm D1})} \rangle$ is found to monotonically decrease from 30.7 Hz [Fig.~\ref{fig:PD}(b)].
Thus, D1 SPNs are under-active due to loss of DA, leading to occurrence of under-active DP.

In the case of D2 SPNs, Figs.~~\ref{fig:PD}(c1)-\ref{fig:PD}(c4) show raster plots of spikes and IPSRs for $x_{DA}$= 0.9, 0.6, 0.4, and 0.1, respectively.
In contrast to the case of D1 SPNs, their (population-averaged) MFR $\langle f_i^{({\rm D2})} \rangle$ is found to monotonically increase from 24.1 Hz [Fig.~\ref{fig:PD}(d)]. Thus, D2 SPNs are over-active because of loss of DA, resulting in appearance of over-active IP.

In the case of STN and GP, associated with IP, their population firing behaviors are shown in their raster plots of spikes and IPSRs for $x_{DA}=$ 0.9, 0.6, 0.4, and 0.1 [see Figs.~\ref{fig:PD}(e1)-\ref{fig:PD}(e4) for STN and see Figs.~\ref{fig:PD}(g1)-\ref{fig:PD}(g4) for GP].
Due to over-active firing activity of the D2 SPNs, the (population-averaged) MFR $\langle f_i^{({\rm GP})} \rangle$ of GP cells is found to monotonically decrease with $x_{DA}$ from 7.3 Hz [Fig.~\ref{fig:PD}(h)]. Also, because of reduced firing activity of the GP cells, the (population-averaged) MFR $\langle f_i^{({\rm STN})} \rangle$ of STN cells is found to monotonically increase with $x_{DA}$ from 39.8 Hz [Fig.~\ref{fig:PD}(f)].

Figure \ref{fig:PD}(i) shows the plot of strengths of DP (green) and IP (red), ${\cal S}_{DP}$ and ${\cal S}_{IP}$, versus $x_{DA}$.
We note that, with decreasing $x_{DA}$ from 1, ${\cal S}_{IP}$ increases rapidly (i.e., over-active IP), while ${\cal S}_{dP}$ decreases very slowly
(i.e., under-active DP). Then, the competition degree ${\cal C}_d$ between DP and IP, given by the ratio of ${\cal S}_{IP}$ to ${\cal S}_{DP}$,
is found to monotonically decrease from ${\cal C}_d^*$ (=2.82), corresponding to that in the healthy state with harmony between DP and IP).
When passing a threshold $x_{DA}^*~ (\simeq 0.27$), ${\cal C}_d=1$ (i.e., DP and IP are balanced); for $x_{DA} > x_{DA}^*,$  ${\cal C}_d >1,$ while for
for $x_{DA} < x_{DA}^*,$  ${\cal C}_d < 1.$

Figures \ref{fig:PD}(k1)-\ref{fig:PD}(k4) and \ref{fig:PD}(l) show population and individual firing behaviors of the output SNr cells, respectively.
With decreasing $x_{DA}$ from 1, their population-averaged MFR $\langle f_i^{({\rm SNr})} \rangle$ is found to monotonically increase from
5.5 Hz (corresponding to that in the healthy state). When $x_{DA}$ passes its threshold, $x_{DA}^*$ ($\simeq 0.27$), $\langle f_i^{({\rm SNr})} \rangle$ becomes
larger than 25.5 Hz [corresponding to that in the default state with ${\cal C}_d \simeq 1$, and represented by the horizontal dashed line in Fig.~\ref{fig:PD}(l)].

Due to loss of DA ($x_{DA} < 1$), IP becomes highly over-active, while DP becomes under-active, in comparison to the healthy state with $x_{DA}=1$.
For $1 > x_{DA} > x_{DA}^*~(\simeq~0.27)$, ${\cal C}_d^* (=2.82)  > {\cal C}_d > 1$. In this case, DP is still stronger than IP, and hence the BG gate to the thalamus is opened. But, the (population-averaged) MFR of SNr cells, $\langle f_i^{({\rm SNr})} \rangle$, is larger than that (= 5.5 Hz) for the healthy state with ${\cal C}_d^*~(=2.82)$. Hence, with decreasing $x_{DA}$ from 1, the ``opening'' degree (of the BG gate to the thalamus) is gradually reduced (i.e., occurrence of break-up of harmony between DP and IP), resulting in appearance of a pathological Parkinsonian state (e.g., PD showing abnormal impaired movement)) with disharmony between DP and IP. For $x_{DA} < x_{DA}^*,$ ${\cal C}_d < 1$ and $\langle f_i^{({\rm SNr})} \rangle > 25.5$ Hz. In this case, IP is stronger than DP, and hence the BG gate to the thalamus becomes locked, leading to no movement. As $x_{DA}$ is decreased from $x_{DA}^*$ the ``locking'' degree of the BG gate (to the thalamus) is increased.

\subsection{Treatment of Pathological Parkinsonian States via Recovery of Harmony between DP and IP}
\label{subsec:Treat}
For the pathological Parkinsonian state, IP is over-active, while DP is under-active, in comparison to the healthy state.
In this way, harmony between DP and IP is broken up in the case of the pathological Parkinsonian state (i.e. occurrence of disharmony between DP and IP).
Here, we investigate treatment of the pathological state with reduced competition degree ${\cal C}_d$ [$< {\cal C}_d^*$ (= 2.82 for the healthy state)]
via recovery of harmony between DP and IP.

In Fig.~\ref{fig:Acti}, activation and deactivation of the target cells via optogenetics are studied. When the light-sensitive proteins (called the opsins)
are activated by specific light stimulation, variation in the intrinsic ionic currents of the cells in the target population $X$, $\Delta I_{ion}^{(X)}$, occurs.
When $\Delta I_{ion}^{(X)}$ is positive (negative), firing activity of the target cells is increased (decreased), resulting in their activation (deactivation)
\cite{OG1,OG2}. As discussed there, we simulate the effects of optogenetics by including $\Delta I_{ion}^{(X)}$ in Eq.~(\ref{eq:GE1}) (in Appendix \ref{app:A}),
in addition to the current, $I_i^{(X)}$, into the target $X$ population. As the intensity of light stimulation is increased, the magnitude of $\Delta I_{ion}^{(X)}$
also increases.

As an example, we consider the pathological state with ${\cal C}_d=1.71$ for $x_{DA}=0.6$ where harmony between DP and IP is broken up.
In this pathological state, DP is under-active. Hence, we first strengthen the DP via activation of the target D1 SPNs. Figure \ref{fig:Treat}(a1) shows plots of ${\cal S}_{DP}$ (strength of DP) and ${\cal S}_{IP}$ (strength of IP) versus $\Delta I_{ion}^{({\rm D1})}$. ${\cal S}_{DP}$ (green) increases rapidly from 2200, while ${\cal S}_{IP}$ (red) remains unchanged (i.e., 1288.9). Thanks to the strengthened DP, the competition degree ${\cal C}_d$ between DP and IP is found to increase from 1.71 [Fig.~\ref{fig:Treat}(a2)]. Also, the population-averaged MFR of the output SNr cells, $\langle f_i^{({\rm SNr})} \rangle$, is found to decrease from 13 Hz [Fig.~\ref{fig:Treat}(a3)].

We note that, when $\Delta I_{ion}^{({\rm D1})}$ passes a threshold $\Delta I_{ion}^{({\rm D1})*}$ (= 51 pA), ${\cal C}_d~=~ {\cal C}_d^*~(=~2.82)$ and
$\langle f_i^{({\rm SNr})} \rangle~ =~ \langle f_i^{({\rm SNr})*} \rangle$ (= 5.5 Hz); ${\cal C}_d^*$ and $\langle f_i^{({\rm SNr})*} \rangle$ are those for the healthy state, and they are represented by the horizontal dashed lines in Figs.~\ref{fig:Treat}(a2) and \ref{fig:Treat}(a3).
Thus, for $x_{DA}=0.6$, the pathological state with ${\cal C}_d = 1.71$ may have ${\cal C}_d^*$ (= 2.82) via activation of D1 SPNs for
the threshold, $\Delta I_{ion}^{({\rm D1})*}$ (= 51 pA); DP becomes 2.82 times stronger than IP, as in the case of healthy state.
In this way, harmony between DP and IP is recovered for $\Delta I_{ion}^{({\rm D1})*}$ = 51 pA.
Figure \ref{fig:Treat}(b) shows the plot of $\Delta I_{ion}^{({\rm D1})*}$ versus $x_{DA}$.
As $x_{DA}$ is decreased from 1, the threshold $\Delta I_{ion}^{({\rm D1})*}$ is increased;
with decreasing $x_{DA}$, more $\Delta I_{ion}^{({\rm D1})*}$ is necessary for recovery of harmony between DP and IP.

In the pathological state for $x_{DA}=0.6$, IP is over-active. Hence, for recovery of harmony between DP and IP, we try to weaken the IP via deactivation of D2 SPNs or STN cells; in the case of deactivation, $\Delta I_{ion}^{(X)}$ [$X=$ D2 (SPN) and STN] is negative, in contrast to the case of activation with
$\Delta I_{ion}^{({\rm D1})} > 0$.
Figures \ref{fig:Treat}(c1)- \ref{fig:Treat}(c3) and \ref{fig:Treat}(d) show the case of deactivation of D2 SPNs.
As the magnitude of $\Delta I_{ion}^{({\rm D2})}$ is increased (i.e., more negative), strength of IP, ${\cal S}_{IP}$ (red), is found to decrease from 1288.9,  while
${\cal S}_{DP}$ (green) remains constant (= 2200). Due to the weakened IP, the competition degree ${\cal C}_d$ between DP and IP increases from 1.71 [Fig.~\ref{fig:Treat}(c2)], and the population-averaged MFR of the output SNr cells, $\langle f_i^{({\rm SNr})} \rangle$, decreases from 13 Hz [Fig.~\ref{fig:Treat}(c3)]. When passing a threshold $\Delta I_{ion}^{({\rm D2})*}$ (= -65 pA), the competition degree ${\cal C}_d$ and the population-averaged
MFR $\langle f_i^{({\rm SNr})} \rangle$ recover their values for the healthy state,  ${\cal C}_d^*$ (= 2.82) and $\langle f_i^{({\rm SNr})*} \rangle$ (= 5.5 Hz),
as in the above case of activation of D1 SPNs. Thus, balance between DP and IP becomes recovered for $\Delta I_{ion}^{({\rm D2})*}$ = -65 pA.
Figure \ref{fig:Treat}(d) shows the plot of $\Delta I_{ion}^{({\rm D2})*}$ versus $x_{DA}$.
With decreasing $x_{DA}$ from 1, the threshold $\Delta I_{ion}^{({\rm D2})*}$ is decreased (i.e., its magnitude increases).
As $x_{DA}$ is decreased from 1, more negative $\Delta I_{ion}^{({\rm D2})*}$ is required for recovery of harmony between DP and IP.

We also study the case of deactivation of STN to weaken the IP.
Figures \ref{fig:Treat}(e1)- \ref{fig:Treat}(e3) and \ref{fig:Treat}(f) show the case of deactivation of STN.
We note that the process of deactivation for STN is similar to that for D2 SPNs.
Thus, when $\Delta I_{ion}^{({\rm STN})}$ passes a threshold, $\Delta I_{ion}^{({\rm STN})*}$ (= -42 pA), harmony between DP and IP becomes
recovered (i.e., ${\cal C}_d$ and $\langle f_i^{({\rm SNr})} \rangle$ have their values for the healthy state) [Figs.~\ref{fig:Treat}(e2) and \ref{fig:Treat}(e3)].
As $x_{DA}$ is decreased from 1, the threshold value of $\Delta I_{ion}^{({\rm STN})*}$ is found to decrease, and hence more negative
$\Delta I_{ion}^{({\rm STN})*}$ is necessary to get recovery of harmony between DP and IP [Fig.~\ref{fig:Treat}(f)].

Finally, instead of the above activation/deactivation via optogenetics, we also consider ablation of STN cells in the pathological state for $x_{DA}=0.6$
to reduce the over-activity of STN cells. In the case of ablation, the number of STN cells, $N_{{\rm STN}}$, is reduced to $N_{\rm STN}^{(n)}$ $x_{\rm STN}$
($ 1 > x_{\rm STN} \geq 0)$, where $N_{\rm STN}^{(n)}$ (= 14) is the normal number of STN cells and $x_{\rm STN}$ is the fraction of number of STN cells.
We note that, the effect of decreasing $x_{\rm STN}$ via ablation is similar to that of deactivation of STN cells via optogenetics.
Figures \ref{fig:Treat}(g1)- \ref{fig:Treat}(g3) and \ref{fig:Treat}(h) show the case of ablation of STN cells.
With decreasing $x_{\rm STN}$ from 1, strength of IP, ${\cal S}_{IP}$ (red), is found to decrease from 1288.9 (i.e., IP becomes weakened)
[Fig.~\ref{fig:Treat}(g1)]. When passing a threshold, $x_{\rm STN}^*~(\simeq 0.51)$, harmony between DP and IP becomes recovered;
${\cal C}_d$ and $\langle f_i^{({\rm SNr})} \rangle$ have their values for the healthy state with harmony between DP and IP [Figs.~\ref{fig:Treat}(g2) and
\ref{fig:Treat}(g3)]. Figure \ref{fig:Treat}(h) shows the plot of $x_{\rm STN}^*$ versus $x_{DA}$.
As $x_{DA}$ is decreased, $x_{\rm STN}^*$ decreases; more ablation (i.e., smaller $x_{\rm STN}$) is necessary for harmony between DP and IP.

\section{Summary and Discussion}
\label{sec:SUM}
The BG exhibit diverse functions for motor and cognition. They control voluntary movement and also make a crucial role in cognitive processes (e.g., action selection). Dysfunction in the BG is related to movement disorder (e.g., PD) and cognitive disorder.
Two competitive pathways exist in the BG; ``Go'' DP (which facilitates movement) and ``No-Go'' IP (which suppresses movement) \cite{DIP1,DIP2,DIP3,DIP4,Frank1,Frank2,Go1,Go2}.
A variety of functions of the BG have been known to be done via ``balance'' between DP and IP \cite{Luo,Kandel,Squire,Bear}.
However, so far, to the best of our knowledge, no quantitative analysis for such balance was made.

In this paper, to make clear the concept of such traditional ``balance,'' as a first time, we quantified competitive harmony (i.e., competition and cooperative interplay) between ``Go'' DP and ``No-Go'' IP by introducing their competition degree ${\cal C}_d$, provided by the ratio of strength of DP (${\cal S}_{DP})$ to strength of IP (${\cal S}_{IP})$; ${\cal C}_d = {\cal S}_{DP} / {\cal S}_{IP}$. Here, ${\cal S}_{DP}$ (${\cal S}_{IP}$) is just the magnitude of time-averaged DP (IP) presynaptic current into the SNr (output nucleus); ${\cal S}_{DP}$ (${\cal S}_{IP}$) = $| \overline {I_{DP}(t)} |$ ($| \overline { I_{IP}(t)} |$) (the overline represents time averaging). These newly-introduced ${\cal C}_d$ was found to well characterize competitive harmony between DP and IP.

The case of normal DA level of $\phi^*=0.3$ was first considered.
A default BG state with ${\cal C}_d \simeq 1$ [i.e., DP and IP are nearly balanced (i.e., nearly equal weighted)] was found to appear for the tonic cortical input (3 Hz) in the resting state. In this default case, the firing activities of the output SNr cells are very active with the firing frequency $f=$ 25.5 Hz, leading to the
locked state of the BG gate to the thalamus. As a result, no voluntary movement occurs.
In contrast, for the phasic cortical input (10 Hz) in the phasically-active state, a healthy state with ${\cal C}_d^* = 2.82$ was found to appear.
In this healthy case, DP is 2.82 times stronger than IP, in contrast to the default case with balanced DP and IP.
Due to more activeness of DP, the firing frequency of the SNr cells becomes much reduced to 5.5 Hz, resulting in the opened state of the BG gate to the
thalamus. Consequently, for the healthy state with ${\cal C}_d^* = 2.82$, normal movement occurs via competitive harmony between DP and IP.

However, as the DA level, $\phi = \phi^*(=0.3)~x_{DA}$ ($1 > x_{DA} \geq 0$), is reduced [i.e., with decreasing $x_{DA}$ (corresponding to fraction of the DA level) from 1], the competition degree ${\cal C}_d$ between DP and IP was found to monotonically decrease from ${\cal C}_d^*$, resulting in appearance of a pathological Parkinsonian state. In the case of the pathological Parkinsonian state, strength of IP (${\cal S}_{IP}$) was found to be much increased than that for the normal healthy state, which leads to  disharmony between DP and IP. Due to break-up of harmony between DP and IP, generating from deficiency in DA production in the cells of the SNc \cite{PD5,PD6}, a pathological Parkinsonian state (e.g., PD with impaired movement) with reduced ${\cal C}_d$ occurs.

In the case of the pathological Parkinsonian state (i.e., PD), DP is under-active, while IP is over-active, in comparison to the healthy state.
In this case, we also investigated treatment of the pathological Parkinsonian state via recovery of harmony between DP and IP.
We included the effects of optogenetics \cite{OG1,OG2}, activating/deactivating the target cells, in the governing equations of their states by adding
$\Delta I_{ion}^{(X)}$ (variation in the intrinsic ionic current of the target cells caused by the optogenetics).
DP was found to be strengthened via activation of D1 SPNs, while IP was found to be weakened through deactivation of D2 SPNs or STN cells.
As a result of this kind of activation/deactivation, the competition degree (${\cal C}_d$) and the population-averaged MFR
($\langle f_i^{({\rm SNr})} \rangle$) of the SNr cells were found to have their ones for the healthy state, [i.e., ${\cal C}_d^* =2.82$ and
$\langle f_i^{({\rm SNr})*} \rangle=$ 5.5 Hz]. In this way, treatment was done through recovery of harmony between DP and IP.

As explained in the above cases of healthy and pathological Parkinsonian states, the newly-introduced competition degree ${\cal C}_d$ between ``Go'' DP and ``No-Go'' IP was found to play an important role of quantifying harmony between DP and IP. Through this kind of quantitative analysis of their harmony, our understanding of their traditional ``balance'' could be quantitatively and clearly improved. Hence, we expect that in future, ${\cal C}_d$ could also be employed for quantitative analysis of harmony between DP and IP in experimental and clinical works. We briefly suggest a possibility of getting ${\cal C}_d$ experimentally.
By using the techniques (e.g., triple patch recording \cite{Exp1} and optical measurement \cite{Exp2}) to measure the presynaptic currents, both the DP and IP presynaptic currents, $I_{DP}(t)$ and $I_{IP}(t),$ into the output nucleus SNr could be measured in experimental works, and then the strengths of the DP and the IP, ${\cal S}_{DP}$ and ${\cal S}_{IP}$, could be obtained via time-average of $I_{DP}(t)$ and $I_{IP}(t),$ respectively. Thus, the competition degree ${\cal C}_d$ between the DP and the IP (given by the ratio of ${\cal S}_{DP}$  to ${\cal S}_{IP}$) could be got experimentally.
In this way, via direct experimental measurement of $I_{DP}(t)$ and $I_{IP}(t),$ quantifying harmony between DP and IP in experimental works could also be possible, which would lead to improving our understanding of harmony between DP and IP much clearly.

Finally, we discuss limitations of our present work and future works.
In addition to motor control, the BG plays an important role in cognitive processes such as action selection \cite{GPR1,GPR2,Hump1,Hump2,Hump3,Man,CN14}.
In this case, a BG network with parallel channels, representing different action requests, arising from the cortex, is usually considered.
Saliency of a channel may be given by the firing frequency of its cortical input; the higher frequency denotes the higher saliency.
Resolution of competition between the channels may be given by selection of a desired channel with the highest salience.
Firing activities of the SNr cells in the highest salient channel are suppressed below a threshold, and hence action in this
channel is selected. On the other hand, in the other neighboring channels, firing activities of the SNr cells are enhanced above the
threshold, and hence actions in these channels are not selected. As a future work, we could apply our present approach, based on the competition degree
${\cal C}_d$, to the case of action selection. Saliency of each channel may be given by its ${\cal C}_d$. Then, action in the channel with the highest
${\cal C}_d$ could be selected.

Next, in future, we would like to consider more realistic SNN for the BG.
In the present SNN, we consider only the D1/D2 SPNs (95 $\%$ major population) in the striatum (primary input nucleus in BG).
However, the remaining minor population of fast-spiking interneurons (FSIs) are found to exert strong influences on spiking activities of the D1/D2 SPNs \cite{Str2,FSI}.
Thus, it is worth while to include the FSIs in the SNN for the BG. Of course, the effects of DA on the FSIs and their synaptic inputs must also be considered.
In this way, to take into consideration the influences of the FSIs would be suitable for more complete SNN for the BG.
Moreover, it would be desirable that, the present BG SNN with cortical inputs modelled in terms of Poisson spike trains is enlarged to the cortico-BG-thalamo-cortical (CBGTC) loop through inclusion of the cortical and the thalamic cells \cite{CN1,CBGTC}.

We also discuss application of the optogenetic techniques to human patients for treatment of PD \cite{OG3,OG4}.
In a pathological Parkinsonian state (i.e., PD) with reduced competition degree ${\cal C}_d$, harmony between DP and IP is broken up; DP is under-active, while
IP is over-active, in comparison to the healthy case with ${\cal C}_d^*~(=2.82)$. As in Sec.~\ref{subsec:Treat}, such harmony between DP and IP could be
recovered by strengthening DP or weakening IP. To this aim, optogenetics could be used. Activation of D1 SPNs via optogenetics
leads to strengthening DP and deactivation of D2 SPNs or STN cells through optogenetics results in weakening IP.
Thus, the competition degree ${\cal C}_d$ between DP and IP becomes increased to ${\cal C}_d^*~(=2.82)$ (i.e., their harmony becomes recovered).
We hope that, in near future, safe clinical applications of optogenetic techniques to human patients with PD could be effectively available.
Then, a substantial step forward for treatment of PD would be taken.

\section*{Acknowledgments}
This research was supported by the Basic Science Research Program through the National Research Foundation of Korea (NRF) funded by the Ministry of Education (Grant No. 20162007688).

\appendix
\section{Single Neuron Models and DA Effects}
\label{app:A}
\begin{table*}
\caption{9 single-cell parameter values in the $X$ (= D1 SPN, D2 SPN, STN, GP, SNr) population.}
\label{tab:Single}
\begin{tabular}{|c|c|c|c|c|}
\hline
Parmeters & \hspace{0.5cm} D1/D2 SPN \hspace{0.5cm} & \hspace{1cm} STN \hspace{1cm} & \hspace{1cm} GP \hspace{1cm} & \hspace{1cm} SNr \hspace{1cm} \\
\hline
$C_X$ & 16.1 & 23.0 & 68.0 & 172.1 \\
\hline
$v_r^{(X)}$ & -80.0 & -56.2 & -53.0 & -64.58 \\
\hline
$v_t^{(X)}$ & -29.3 & -41.4 & -44.0 & -51.8 \\
\hline
$k_X$ & 1 & 0.439 & 0.943 & 0.7836 \\
\hline
$a_X$ & 0.01 & 0.021 & 0.0045 & 0.113 \\
\hline
$b_X$ & -20 & 4 & 3.895 & 11.057 \\
\hline
$c_X$ & -55 & -47.7 & -58.36 & -62.7 \\
\hline
$d_X$ & 84.2 & 17.1 & 0.353 & 138.4 \\
\hline
$v_{peak}^{(X)}$ & 40 & 15.4 & 25 & 9.8 \\
\hline
\end{tabular}
\end{table*}
The Izhikevich neuron models are considered as single neuron models in the BG SNN  \cite{Izhi1,Izhi2,Izhi3,Izhi4}.
Evolution of dynamical states of individual cells in the $X$ population [$X=$ D1 (SPN), D2 (SPN), STN, GP, and SNr]
is governed by the following equations:
\begin{eqnarray}
C_X \frac{dv_i^{(X)}}{dt} &=& k_X (v_i^{(X)} - v_r^{(X)}) (v_i^{(X)} - v_t^{(X)}) \nonumber \\
& & - u_i^{(X)} +I_i^{(X)}, \label{eq:GE1} \\
\frac{du_i^{(X)}}{dt} &=& a_X \left\{b_X (v_i^{(X)} - v_r^{(X)}) - u_i^{(X)} \right\}; \nonumber \\
& & ~~~~~~ i = 1, ..., N_X, \label{eq:GE2}
\end{eqnarray}
with the auxiliary after-spike resetting:
\begin{equation}
{\rm if}~ v_i^{(X)} \ge v_{peak}^{(X)}, ~{\rm then}~ v_i^{(X)} \leftarrow c_X ~{\rm and}~ u_i^{(X)} \leftarrow u_i^{(X)}+d_X ,
\label{eq:Reset}
\end{equation}
where $N_X$ and $I_i^{(X)}(t)$ are the total number of cells and the current into the $i$th cell in the $X$ population, respectively.
In Eqs.~(\ref{eq:GE1}) and (\ref{eq:GE2}), the dynamical state of the $i$th cell in the $X$ population at a time $t$ (msec) is characterized by its membrane potential $v_i^{(X)}(t)$ (mV) and the slow recovery variable $u_i^{(X)}(t)$ (pA).
When $v_i^{(X)}(t)$ reaches a threshold $v_{peak}^{(X)}$ (i.e., spike cutoff value), firing a spike occurs, and then
$v_i^{(X)}$ and $u_i^{(X)}$ are reset in accord with the rules of Eq.~(\ref{eq:Reset}).

There are 9 intrinsic parameters in each $X$ population;
$C_X$ (pF): membrane capacitance, $v_r^{(X)}$ (mV): resting membrane potential, $v_t^{(X)}$ (mV): instantaneous threshold potential,
$k_X$ (nS/mV): parameter associated with the cell’s rheobase, $a_X$ (${\rm msec}^{-1}$): recovery time constant,
$b_X$ (nS): parameter associated with the input resistance, $c_X$ (mV): after-spike reset value of $v_i^{(X)}$,
$d_X$ (pA): after-spike jump value of $u_i^{(X)}$, and $v_{peak}^{(X)}$ (mV): spike cutoff value.
Table \ref{tab:Single} shows the 9 intrinsic parameter values of the BG cells.
Along with the parameter values of the D1/D2 SPNs provided in \cite{SPN1,SPN2}, we get the parameter values of the other cells (STN, GP, SNr),
founded on the work in \cite{CN6}. In the case of GP and STN, we consider the major subpopulations of high frequency pauser (85 $\%$) and short rebound bursts
(60 $\%$), respectively. Also, we use the standard 2-variable Izhikevich neuron model for the STN, instead of the 3-variable Izhikevich neuron model in \cite{CN6};
these two models give nearly the same results for the STN.

We also consider influences of DA modulation on the D1 and D2 SPNs \cite{SPN1,SPN2,CN6}.
D1 receptors activation has two opposing influences on intrinsic ion channels.
It enhances the inward-rectifying potassium current (KIR), leading to hyperpolarization of the D1 SPN.
In contrast, it lowers the activation threshold of the L type ${\rm Ca}^{2+}$ current, resulting in depolarization of the D1 SPN.
These two hyperpolarization and depolarization influences are modelled via variations in intrinsic parameters of the D1 SPN:
\begin{eqnarray}
v_r &\leftarrow& v_r (1+\beta_1^{({\rm D1})} \phi_1), \label{eq:DA-D1v} \\
d &\leftarrow& d(1-\beta_2^{({\rm D1})} \phi_1). \label{eq:DA-D1d}
\end{eqnarray}
Here, Eq.~(\ref{eq:DA-D1v}) models the hyperpolarizing effect of the increasing KIR by upscaling $v_r$, while Eq.~(\ref{eq:DA-D1d}) models enhanced depolarizing effect of the L type ${\rm Ca}^{2+}$ current by downscaling $d$. The parameters $\beta_1^{(D1)}$ and $\beta_2^{(D1)}$ represent the amplitudes of their respective influences, and $\phi_1$ is the DA level (i.e., fraction of active DA receptors) for the D1 SPNs.

\begin{table}
\caption{Effects of DA modulation on intrinsic parameters of the D1/D2 SPNs.}
\label{tab:DASPN}
\begin{tabular}{|c|c|c|}
\hline
\multirow{2}{*}{D1 SPN} & $v_r \leftarrow v_r (1+\beta_1^{({\rm D1})} \phi_1)$ & $\beta_1^{({\rm D1})}=0.0289$ \\
\cline{2-3}
 & $d \leftarrow d(1-\beta_2^{({\rm D1})} \phi_1)$ & $\beta_2^{({\rm D1})}=0.331$ \\
\hline
D2 SPN & $k \leftarrow k (1-\beta_1^{({\rm D2})} \phi_2)$ & $\beta^{({\rm D2})}=0.032$ \\
\hline
\end{tabular}
\end{table}

Next, D2 receptors activation has small inhibitory influence on the slow A-type potassium current, leading to decrease in
the cell's rheobase current. This depolarizing effect is well modelled by downscaling the parameter, $k$:
\begin{equation}
k \leftarrow k (1-\beta^{({\rm D2})} \phi_2),
\label{DA-D2}
\end{equation}
where $\phi_2$ is the DA level for the D2 SPNs, and the parameter $\beta^{(D2)}$ represents the downscaling degree in $k$.
Table \ref{tab:DASPN} shows DA modulation on the intrinsic parameters of the D1/D2 SPNs where the parameter values of
$\beta_1^{(D1)}$, $\beta_2^{(D1)}$, and $\beta^{(D2)}$ are given \cite{SPN1,SPN2,CN6}. In this paper, we consider the case of $\phi_1 = \phi_2 = \phi$.

Time-evolution of $v_i^{(X)}(t)$ and $u_i^{(X)}(t)$ in Eqs.~(\ref{eq:GE1}) and (\ref{eq:GE2}) is governed by the current $I_i^{(X)}(t)$ into the $i$th cell in the $X$ population:
\begin{equation}
I_i^{(X)}(t) = I_{ext,i}^{(X)}(t) - I_{syn,i}^{(X)}(t) + I_{stim}^{(X)}(t).
\label{eq:I}
\end{equation}
Here, $I_{ext,i}^{(X)}$, $I_{syn,i}^{(X)}(t)$, and $I_{stim}^{(X)}(t)$ denote the external current from the external background region (which is not considered in the modeling), the synaptic current, and the injected stimulation current, respectively. In the BG SNN, we consider the case of $I_{stim}=0$ (i.e., no injected stimulation DC current).

\begin{table}
\caption{In-vivo firing activities of BG cells in awake resting state with tonic cortical input (3 Hz) in the case of the normal DA level of $\phi=0.3$.
Spontaneous current $I_{vivo}^{(X)}$, firing rates $f_{vivo}^{(X)}$, and random background input $D_X^*$ ($X=$ D1 SPN, D2 SPN, STN, GP, and SNr)}
\label{tab:Spon}
\begin{tabular}{|c|c|c|c|c|}
\hline
Parmeters & D1/D2 SPN & \hspace{0.3cm} STN \hspace{0.3cm} & \hspace{0.3cm} GP \hspace{0.3cm} & \hspace{0.3cm} SNr \hspace{0.3cm} \\ \hline
$I_{vivo}^{(X)}$ & 0 & 56.5 & 84.0 & 292.0 \\
\hline
$f_{vivo}^{(X)}$ & 1 & 9.9 & 29.9 & 25.5 \\
\hline
$D_X^*$ & 246 & 11.9 & 274 & 942 \\
\hline
\end{tabular}
\end{table}

The external current $I_{ext,i}^{(X)}(t)$ may be modeled in terms of $I_{spon,i}^{(X)}$ [spontaneous current for spontaneous firing activity, corresponding to time average of $I_{ext,i}^{(X)}(t)$] and $I_{back,i}^{(X)}(t)$ [random background input, corresponding to fluctuation from time average of $I_{ext,i}^{(X)}(t)$].
In the BG population, $I_{spon}^{(X)}$ (independent of $i$) is just the spontaneous in-vivo current, $I_{vivo}^{(X)}$, to get the spontaneous in-vivo firing rate $f_{vivo}^{(X)}$ in the presence of synaptic inputs in the resting state (in-vivo recording in awake resting state with tonic cortical input).
The random background current $I_{back,i}^{(X)}(t)$ is given by:
\begin{equation}
I_{back,i}^{(X)}(t) = D_X \cdot  \xi_i^{(X)}(t).
\label{eq:Iback}
\end{equation}
Here, $D_X$ is the parameter controlling the noise intensity and $\xi_i^{(X)}$ is the Gaussian white noise, satisfying the zero mean and the unit variance
\cite{Kim1,Kim2,Kim3}:
\begin{equation}
\langle \xi_i^{(X)}(t) \rangle = 0 ~{\rm and}~ \langle \xi_i^{(X)}(t) \xi_j^{(X)}(t') \rangle = \delta_{ij}\delta(t-t').
\label{eq:Noise}
\end{equation}
Table~\ref{tab:Spon} shows in-vivo firing activities of BG cells in awake resting state with tonic cortical input for the normal DA level of $\phi=0.3$;
spontaneous in-vivo currents $I_{vivo}^{(X)}$, in-vivo firing rates $f_{vivo}^{(X)}$, and random background inputs $D_X^*$ for  \cite{Hump1,CN6,CN11} are given.

\section{Synaptic Currents and DA Effects}
\label{app:B}
We explain the synaptic current $I_{syn,i}^{(X)}(t)$ in Eq.~(\ref{eq:I}).
There are two kinds of excitatory synaptic currents, $I_{{\rm AMPA},i}^{(X,Y)}(t)$ and $I_{{\rm NMDA},i}^{(X,Y)}(t)$, which are
are the AMPA ($\alpha$-amino-3-hydroxy-5-methyl-4-isoxazolepropionic acid) receptor-mediated and NMDA ($N$-methyl-$D$-aspartate) receptor-mediated currents from the presynaptic source $Y$ population to the postsynaptic $i$th cell in the target $X$ population, respectively.
In addition to these excitatory synaptic currents, there exists another inhibitory synaptic current, $I_{{\rm GABA},i}^{(X,Z)}(t)$,
which is the $\rm GABA_A$ ($\gamma$-aminobutyric acid type A) receptor-mediated current from the presynaptic source $Z$ population to the postsynaptic $i$th cell in the target $X$ population.

Here, we follow the ``canonical'' formalism for the synaptic currents, as in our previous works in the cerebellum \cite{Cere1,Cere2} and the hippocampus
\cite{Hippo1,Hippo2,Hippo3,Hippo4}. The synaptic current $I_{R,i}^{(T,S)}(t)$ $R$ (= AMPA, NMDA, or GABA) obeys the following equation:
\begin{equation}
I_{R,i}^{(T,S)}(t) = g_{R,i}^{(T,S)}(t)~(v_{i}^{(T)}(t) - V_{R}^{(S)}),
\label{eq:ISyn1}
\end{equation}
where $g_{(R,i)}^{(T,S)}(t)$ and $V_R^{(S)}$ are synaptic conductance and synaptic reversal potential, respectively.

The synaptic conductance $g_{R,i}^{(T,S)}(t)$ is given by:
\begin{equation}
g_{R,i}^{(T,S)}(t) = {\widetilde g}_{max,R}^{(T,S)} \sum_{j=1}^{N_S} w_{ij}^{(T,S)} ~ s_{j}^{(T,S)}(t),
\label{eq:ISyn2}
\end{equation}
where ${\widetilde g}_{max,R}^{(T,S)}$ and $N_S$ are the maximum synaptic conductance and the number of cells in the source population $S$, respectively.
Here, the connection weight $w_{ij}^{(T,S)}$ is 1 when the $j$th presynaptic cell is connected to the $i$th postsynaptic cell;
otherwise (i.e., in the absence of such synaptic connection), $w_{ij}^{(T,S)}=0$ .

\begin{table}
\caption{Synaptic parameter values. $S$: source population, $T$: target population, $R$: receptor, $\tilde{g}_{max,R}^{(T,S)}$: maximum synaptic conductances, $\tau_{R,d}^{(T,S)}$: synaptic decay times, $\tau_{R,l}^{(T,S)}$: synaptic delay times, and $V_R^{(S)}$: synaptic reversal potential.}
\label{tab:SynParm}
\begin{tabular}{|c|c|c|c|c|c|}
\hline
$S \rightarrow T$ & $R$ & $\tilde{g}_{max,R}^{(T,S)}$ & $\tau_{R,d}^{(T,S)}$ & $\tau_{R,l}^{(T,S)}$ & $V_R^{(S)}$ \\
\hline
\multirow{2}{*}{Ctx $\rightarrow$ D1/D2 SPN} & AMPA & 0.6 & 6 & 10 & 0 \\
\cline{2-6}
 & NMDA & 0.3  & 160 & 10 & 0 \\
\hline
\multirow{2}{*}{Ctx $\rightarrow$ STN} & AMPA & 0.388 & 2 & 2.5 & 0 \\
\cline{2-6}
 & NMDA & 0.233  & 100 & 2.5 & 0 \\
\hline
D1 SPN $\rightarrow$ SNr & GABA & 4.5 & 5.2 & 4 & -80 \\
\hline
D2 SPN $\rightarrow$ GP & GABA & 3.0 & 6 & 5 & -65 \\
\hline
\multirow{2}{*}{STN $\rightarrow$ GP} & AMPA & 1.29 & 2 & 2 & 0 \\
\cline{2-6}
 & NMDA & 0.4644  & 100 & 2 & 0 \\
\hline
GP $\leftrightarrow$ GP & GABA & 0.765 & 5 & 1 & -65 \\
\hline
GP $\rightarrow$ STN & GABA & 0.518 & 8 & 4 & -84 \\
\hline
\multirow{2}{*}{STN $\rightarrow$ SNr} & AMPA & 12 & 2 & 1.5 & 0 \\
\cline{2-6}
 & NMDA & 5.04  & 100 & 1.5 & 0 \\
\hline
GP $\rightarrow$ SNr & GABA & 73 & 2.1 & 3 & -80 \\
\hline
\end{tabular}
\end{table}

We note that, $s^{(T,S)}(t)$ in Eq.~(\ref{eq:ISyn2}) denotes fraction of open postsynaptic ion channels which are opened through binding of neurotransmitters
(emitted from the source population $S$). A sum of exponential-decay functions $E_{R}^{(T,S)} (t - t_{f}^{(j)}-\tau_{R,l}^{(T,S)})$ provides
time evolution of $s_j^{(T,S)}(t)$ of the $j$th cell in the source $S$ population:
\begin{equation}
s_{j}^{(T,S)}(t) = \sum_{f=1}^{F_{j}^{(S)}} E_{R}^{(T,S)} (t - t_{f}^{(j)}-\tau_{R,l}^{(T,S)}),
\label{eq:ISyn3}
\end{equation}
where
$F_j^{(S)},$ $t_f^{(j)},$ and $\tau_{R,l}^{(T,S)}$ are the total number of spikes and the $f$th spike time of the $j$th cell, and the synaptic latency time constant, respectively.

\begin{table*}
\caption{Effects of DA modulation on synaptic currents into the target population ($T$); $T$: D1 SPN, D2 SPN, STN, and GP.}
\label{tab:DASyn}
\begin{tabular}{|c|c|c|}
\hline
D1 SPN & $I_{AMPA} + f(v) \cdot I_{NMDA} (1+\beta^{({\rm D1})} \phi_1)$ & $\beta^{({\rm D1})}=0.5$ \\
\hline
D2 SPN & $I_{AMPA} (1-\beta^{({\rm D2})} \phi_2) + f(v) \cdot I_{NMDA}$ & $\beta^{({\rm D2})}=0.3$ \\
\hline
STN & $(I_{AMPA} + f(v)\cdot I_{NMDA})(1-\beta_1^{({\rm STN})} \phi_2) + I_{GABA} (1-\beta_2^{({\rm STN})} \phi_2)$ & $\beta_1^{({\rm STN})}=\beta_2^{({\rm STN})}=0.5$ \\
\hline
GP & $(I_{AMPA} + f(v)\cdot I_{NMDA})(1-\beta_1^{({\rm GP})} \phi_2) + I_{GABA} (1-\beta_2^{({\rm GP})} \phi_2)$ & $\beta_1^{({\rm GP})}=\beta_2^{({\rm GP})}=0.5$ \\
\hline
\end{tabular}
\end{table*}

Similar to our previous works in the cerebellum \cite{Cere1,Cere2}, we use the exponential-decay function $E_{R}^{(T,S)} (t)$:
\begin{equation}
E_{R}^{(T,S)}(t) = e^{-t/\tau_{R,d}^{(T,S)}} \cdot \Theta(t),
\label{eq:ISyn4}
\end{equation}
where $\tau_{R,d}^{(T,S)}$ is the synaptic decay time constant and the Heaviside step function satisfies $\Theta(t)=1$ for $t \geq 0$ and 0 for $t <0$.

We also note that, in the case of NMDA-receptor, the positive magnesium ions Mg$^{2+}$ block some of the postsynaptic NMDA channels.
For this case, fraction of non-blocked NMDA channels is given by a sigmoidal function $f(v^{(T)})$ \cite{SPN1,CN6,NMDA},
\begin{equation}
f(v^{(T)}(t)) = \frac{1}{1+0.28 \cdot [{\rm Mg}^{2+}] \cdot e^{-0.062 v^{(T)}(t)}},
\label{eq:Mg}
\end{equation}
where $v^{(T)}$ is the membrane potential of  a cell in the target population $T$ and $[{\rm Mg}^{2+}]$ is the equilibrium concentration of magnesium ions ($[{\rm Mg}^{2+}]$ = 1 mM).
Then, the synaptic current into the $i$th cell in the target $X$ population becomes
\begin{equation}
I_{syn,i}^{(X)}(t) = I_{{\rm AMPA},i}^{(X,Y)}(t) + f(v_i^{(X)}(t)) \cdot I_{{\rm NMDA},i}^{(X,Y)}(t) + I_{{\rm GABA},i}^{(X,Z)}(t).
\label{eq:ISyn5}
\end{equation}
Table \ref{tab:SynParm} shows the synaptic parameters; synaptic parameter values. $S$: source population, $T$: target population, $R$: receptor, $\tilde{g}_{max,R}^{(T,S)}$: maximum synaptic conductances, $\tau_{R,d}^{(T,S)}$: synaptic decay times, $\tau_{R,l}^{(T,S)}$: synaptic delay times, and $V_R^{(S)}$: synaptic reversal potential.

We also take into consideration the influence of DA modulation on the synaptic currents into D1 SPN, D2 SPN, STN, and GP cells in Fig.~\ref{fig:BGN} \cite{SPN1,SPN2,CN6}. For the synaptic currents into the D1 SPNs, influence of DA modulation ma be modeled by upscaling the NMDA receptor-mediated current $I_{\rm NMDA}$ with the factor $\beta^{({\rm D1})}$:
\begin{equation}
I_{\rm NMDA} \leftarrow I_{\rm NMDA} (1+\beta^{({\rm D1})} \phi_1).
\label{DA-D1Syn}
\end{equation}
Here, $\phi_1$ is the DA level for the D1 SPNs. (There is no DA influence on $I_{\rm AMPA}$ for the D1 SPNs.)
On the other hand, for the synaptic currents into the D2 SPNs, influence of DA modulation could be modeled by downscaling the AMPA receptor-mediated current $I_{\rm AMPA}$ with the factor $\beta^{({\rm D2})}$:
\begin{equation}
I_{\rm AMPA} \leftarrow I_{\rm AMPA} (1-\beta^{({\rm D2})} \phi_2).
\label{DA-D2Syn}
\end{equation}
Here, $\phi_2$ is the DA level for the D2 SPNs. (There is no DA influence on $I_{\rm NMDA}$ for the D2 SPNs.)
The scaling factors $\beta^{({\rm D1})}$ and $\beta^{({\rm D2})}$ are given in Table \ref{tab:DASyn}.
Also, effects of DA modulation on synaptic currents into STN cells and GP cells are well given in Table \ref{tab:DASyn}.
In these cases, all excitatory and inhibitory synaptic currents, $I_{\rm AMPA}$, $I_{\rm NMDA}$, and $I_{\rm GABA}$, are downscaled
with their scaling factors, depending on $\phi_2$. Here, $\phi_1 = \phi_2 = \phi$.

\end{document}